\documentclass[conference,letterpaper, 10pt]{IEEEtran}
\usepackage[utf8]{inputenc} 
\usepackage[T1]{fontenc}
\usepackage{url}
\usepackage{ifthen}
\usepackage{cite}
\usepackage[cmex10]{amsmath}

\usepackage[a4paper, total={7.2in, 10.1in}]{geometry}

\usepackage[hidelinks]{hyperref}
\usepackage{amsmath, epsfig, cite}
\usepackage{amsthm}
\usepackage{amsfonts}
\usepackage{graphicx}
\usepackage{soul}
\usepackage{latexsym}
\usepackage{amssymb}
\usepackage{color}
\usepackage{url}
\usepackage{colortbl}
\usepackage{comment}
\usepackage[dvipsnames]{xcolor}
\usepackage{caption}
\usepackage{subcaption}

\usepackage{hyperref}
\usepackage{cleveref}

\usepackage{xfrac}
\usepackage{diagbox}

\DeclareMathAlphabet{\mathbfsl}{OT1}{ppl}{b}{it} 

\newcommand{\E}{\mathbb{E}}

\newcommand{\cC}{{\cal C}}

\newcommand{\cG}{{\cal G}}

\newcommand{\cS}{{\cal S}}

\newcommand{\cU}{{\cal U}}

\newcommand{\cX}{{\cal X}}
\newcommand{\cY}{{\cal Y}}

\newcommand{\bff}{{\boldsymbol f}}

\newcommand{\bfp}{{\boldsymbol p}}

\newcommand{\bfu}{{\boldsymbol u}}
\newcommand{\bfv}{{\boldsymbol v}}

\newcommand{\bfx}{{\boldsymbol x}}
\newcommand{\bfy}{{\boldsymbol y}}

\theoremstyle{definition}
\newtheorem{theorem}{Theorem}
\newtheorem{lemma}{Lemma}
\newtheorem{remark}{Remark}

\newtheorem{corollary}{Corollary}
\newtheorem{definition}{Definition}

\newtheorem{problem}{Problem}
\newtheorem{claim}{Claim}

\newcommand{\abs}[1]{|#1|}

\newcommand{\set}[2]{\left\{#1\;\left|\; #2\right.\right\}}

\newcommand{\Tmax}[2]{T_{\max}^{#1}(#2)}

\title{\textbf{Covering All Bases: The Next Inning in DNA Sequencing Efficiency
}}
\author{%
  \IEEEauthorblockN{Hadas~Abraham}
  \IEEEauthorblockA{The Henry \& Marilyn Taub faculty of Computer Science\\
                    Technion\\
                    Haifa, Israel\\
                    Email: hadasabraham@campus.technion.ac.il}
\and
 \IEEEauthorblockN{Rayn~Gabrys}
  \IEEEauthorblockA{The Henry \& Marilyn Taub faculty of Computer Science\\
                    Technion\\
                    Haifa, Israel\\
                    Email: yaakobi@cs.technion.ac.il}
  \and
 \IEEEauthorblockN{Eitan~Yaakobi}
  \IEEEauthorblockA{The Henry \& Marilyn Taub faculty of Computer Science\\
                    Technion\\
                    Haifa, Israel\\
                    Email: yaakobi@cs.technion.ac.il}
}

\author{\IEEEauthorblockN{\textbf{Hadas Abraham}\IEEEauthorrefmark{2}, \textbf{Ryan Gabrys}\IEEEauthorrefmark{3}, and \textbf{Eitan Yaakobi}\IEEEauthorrefmark{2}}
\vspace{-.29ex}
\IEEEauthorblockA{\IEEEauthorrefmark{2}{\begin{small}The Henry and Marilyn Faculty of Computer Science, Technion -- Israel Institute of Technology, Haifa, Israel.\end{small}}}
\vspace{-.29ex}\IEEEauthorblockA{\IEEEauthorrefmark{3}\begin{small}Calit2, University of California, San Diego.\end{small}}
\vspace{-.29ex} \IEEEauthorblockA { Emails: hadasabraham@campus.technion.ac.il, yaakobi@cs.technion.ac.il, rgabrys@eng.ucsd.edu}
\vspace{-.29ex}}

\IEEEoverridecommandlockouts
\IEEEaftertitletext{\vspace{-3.25ex}}

\IEEEoverridecommandlockouts
\IEEEaftertitletext{\vspace{-3.25ex}}

\begin{document} 

\maketitle

\begin{abstract}
DNA emerges as a promising medium for the exponential growth of digital data due to its density and durability. This study extends recent research by addressing the \emph{coverage depth problem} in practical scenarios, exploring optimal error-correcting code pairings with DNA storage systems to minimize coverage depth. Conducted within random access settings, the study provides theoretical analyses and experimental simulations to examine the expectation and probability distribution of samples needed for files recovery. Structured into sections covering definitions, analyses, lower bounds, and comparative evaluations of coding schemes, the paper unveils insights into effective coding schemes for optimizing DNA storage systems. 
\end{abstract}
\section{Introduction}\label{sec:intro}
\renewcommand{\baselinestretch}{0.955}\normalsize

The rapid growth of digital data, projected to reach 180 zettabytes by 2025, is causing a data storage crisis, with demand surpassing supply\cite{rydning2022worldwide}. Existing storage technologies face challenges meeting big data demands. In response, DNA emerges as a promising medium due to its density and durability. The DNA storage process involves \textit{synthesis}, creating artificial DNA strands encoding user information with limitations leading to short strands and multiple noisy copies \cite{leproust2010synthesis}, storage by a \textit{storage container} and  \textit{sequencing}, a key component \cite{anavy2019data},\cite{erlich2017dna},\cite{organick2018random},\cite{yazdi2017portable}, translates DNA into digital sequences. Despite the potential of DNA storage, current DNA sequencers face challenges such as slow throughput and high costs compared to alternatives\cite{shomorony2022information},\cite{yazdi2015dna},\cite{alliance2021preserving}. Coverage depth, the ratio of sequenced reads to designed strands, impact system latency and costs, highlighting the need for optimization \cite{chandak2019improved},\cite{erlich2017dna}.

We extend recent research addressing the \textit{coverage depth problem} \cite{bar2023cover} by generalizing it to a more practical scenario. Specifically, we consider a container storing $m$ files, each composed of $k$ information strands. These strands are encoded into $mn$ strands using some coding scheme, and the objective is to recover $a$ files out of the total $m$. Our focus is on investigating the required coverage depth, considering factors such as the DNA storage channel and the error-correcting code. Additionally, we aim to explore the optimal pairing of an error-correcting code with a given DNA storage system to minimize coverage depth. This investigation is conducted within the framework of random access settings, where the user seeks to retrieve only a fraction of the stored information. In this context, we conduct both theoretical and experimental analyses to examine the expectation and probability distribution of the number of samples needed to fully recover the specified $a$ files.


The DNA coverage depth problem is akin to well-known problems such as the coupon collector's, dixie cup, and urn problems, where the objective is to collect all types of coupons or objects \cite{erdHos1961classical}, \cite{feller1991introduction}, \cite{flajolet1992birthday}, \cite{newman1960double}. In our context, the "coupons" represent copies of synthesized strands, and the aim is to read at least one copy of each information strand. For example, if $n$ coupons are drawn uniformly at random with repetition, it is well known that the expected number of draws needed to obtain at least one copy of every strand is approximately $n\log n$. However, in this work, we consider the setting where one is allowed to employ the use of a code in order to reduce the number of draws necessary to recover a given subset of information, and it is required to read a specific set of strands that constitute a file.



The paper is structured as follows. In~\autoref{sec:def}, we provide definitions and articulate the problem statement, focusing on the coverage depth problem in our more practical settings. We also discuss some relevant prior results on this matter. In~\autoref{sec:random}, we address the scenario where the user aims to retrieve a single file ($a=1$ out of $m$). We conduct analyses for three coding schemes: the local MDS scheme, which employs an $[n,k]$ MDS code for each of the $m$ files; the global MDS scheme, employing an $[mn,mk]$ systematic MDS code on the combined strands of the $m$ files; and the partial MDS scheme (PMDS), specifically analyzed for the case of $m=2$ files. We present the expected value of samples required to recover a file and explore the expected limit as $n$ approaches infinity for both local and global schemes. In~\autoref{sec:lowerbound}, we establish two lower bounds on the expected number of samples needed for file recovery. \autoref{sec:compare} includes a comparative analysis of the coding schemes. we prove that, in terms of expectation, the local scheme surpasses the global one. Then, a simulation is conducted, providing insights crucial for determining the optimal coding scheme. While the local scheme demonstrates superior expectations, analysis of probability distribution and variance suggests that the global and PMDS schemes may be more favorable options. Finally, in~\autoref{sec:analysisageq1}, we present results for the case where we aim to recover $a \geq 1$ files and extend our lower bound to this case.

\section{Definitions, Problem Statement, Related Work}\label{sec:def}
\subsection{Definitions}
For a positive integer $n$, $[n]$ denotes the set $\{1, \ldots, n\}$ and $H_n$ denotes the $n$-th Harmonic number. We consider a DNA-based storage system in which the data is stored as a codeword, described by a vector of length-$\ell$ sequences or strands over the alphabet $\Sigma = \{A, C, G, T\}$, so the set of all length-$\ell$ vectors over $\Sigma$ is denoted by $\Sigma^\ell$. Often, an outer error-correcting code is employed to protect the data across these length-$\ell$ sequences. In the setting studied in this paper, it is assumed that these strands represent some $m$ files and so the input is represented by a vector of $m$ files $\cU = (U_1, U_2, \ldots, U_m)$, where each file consists of $k$ length-$\ell$ information strands $U_i = (\bfu_{i,1}, \bfu_{i,2}, \ldots, \bfu_{i,k}) \in (\Sigma^\ell)^k$ for $i \in [m]$. The $mk$ information strands are then encoded to $mn$ encoded strands using some linear $[mn,mk]$ code $\cC$ over $\Sigma^\ell$ (typically $\Sigma^\ell$ is embedded into a field of size $4^\ell$). The resulting encoded vector is denoted by $\cX = (\bfx_1, \bfx_2, \ldots, \bfx_{mn})$ which represents the input vector to the DNA storage system. Note that the files can be encoded either seprately or all together.


The DNA storage channel, denoted by $\mathsf{S}$, initially produces numerous noisy copies for each strand in $\cX$. These noisy copies undergo amplification using PCR, and a \emph{sample} of $M$ strands is then sequenced \cite{heckel2019characterization}. The output of the sequencing process is a multiset $\cY_M =\{\!\!\{ \bfy_1, \bfy_2, \ldots, \bfy_M \}\!\!\}$, which consists of \emph{reads} $\bfy_j$ for $j \in [M]$, each being a noisy version of some $\bfx_i$, $i\in[mn]$. The model assumes that the index $i\in[mn]$ such that $\bfy_j$ is a noisy copy of $\bfx_i$ is known. 
The number of reads in $\cY_M$ corresponding to the $i$-th strand $\bfx_i, i\in [mn]$, depends on a categorical probability distribution $\bfp=(p_1,\ldots,p_{mn})$, where for $i\in[mn]$, $p_i$ is the probability to sample a read of $\bfx_i$. However, for simplicity, it is assumed in this work that $\bfp$ is the uniform distribution and we further assume that there is no noise in the reading process so every read in an error-free copy of some $\bfx_i$. Since we consider the noiseless scenario, there is no need to apply clustering or reconstruction algorithm as well as an error-correcting code to correct the errors during reading. However, we do apply an error-correcting in order to reduce the required number of reads in order to decode the information. For a more detailed description of this model which include noise we refer the reader  to~\cite{bar2023cover}. 

\subsection{Problem Statment}
The main goal of this paper is to explore the necessary sample size for the retrieval of some requested $a$ files by the user out of $m$ from $\cU$. Successful decoding of a file $U_i$ for $i\in[m]$ is defined as sampling enough encoded strands from $\cX$ that are sufficient to decode all the $k$ information strands of $U_i$. Note that since the strands in $\cU$ are encoded using an error-correcting code $\cC$, it is not necessary to sample \emph{all} the $k$ information strands from $U_i$ but any set of encoded strands from $\cX$ that allows to decode them. 
We also note that the main difference the model studied in this paper and the one from~\cite{bar2023cover} is that the latter work does not assume the partition of the data into files and considers it as one file. Then, the goal is to either decode the entire file or one information strand. 


Mathematically speaking, assume $\cC$ is the code which is used to encode the $m$ files and let $F\subseteq [m]$ be the set of files that are requested by the user. Let $\nu_{(m,F)}(\cC)$ be the random variable that governs the number of reads that should be sampled for successful decoding of the $a$ files in $F$. The problems studied in this paper are formally defined as follows. 
\begin{problem}\label{prob:random:single}
Given an $[mn,mk]$ code $\cC$,  $F \subseteq [m], |F|=a$. Find the following values:
\begin{enumerate}
\item The expectation value $\E [ \nu_{(m,F)}(\cC)] $ and the probability distribution $\text{Pr}[\nu_{(m,F)}(\cC)>r]$ for any $r\in\mathbb{N}$. 
\item The maximal expected number of samples to retrieve any $a$ files, i.e., 
$$ \Tmax{\cC}{a} \triangleq \max_{|F|=a, F\subseteq [m]} \mathbb{E}[\nu_{(m,F)}(\cC)].$$
\end{enumerate} 
\end{problem}

\begin{problem}\label{prob:optimal}
For given values of $ n, k, m, a$ find:
\begin{enumerate} 
    \item An $[mn,mk]$ code $\cC$,  that is optimal with respect to minimizing $\Tmax{\cC}{a}$. 
    \item The minimum value of $\Tmax{\cC}{a}$ over all possible $[mn,mk]$  codes $\cC$. That is, find the value 
    $ T(n,k;m,a) \triangleq \min_{\cC} \{\Tmax{\cC}{a}\}.$
\end{enumerate}
\end{problem}

In order to address Problem~\ref{prob:random:single}, we consider in Section~\ref{sec:random} three coding schemes and analyze their maximal expected number of samples. These results, in particular, provide an upper bound on the value of $ T(n,k;m,a)$ from Problem~\ref{prob:optimal}, while lower bounds on this value are given in Section~\ref{sec:lowerbound}. 


\subsection{Previous Results}
Two special cases of~\autoref{prob:optimal} have been investigated in~\cite{bar2023cover}. Specifically, in case there is only one file, i.e., $m=1$, which implies that $a=1$ the value of $T(n,k;1,1)$ has been fully solved and it was shown that $T(n,k;1,1) = n(H_n-H_{n-k})$, which is achieved by any $[n,k]$ MDS code. Similarly, it is easily deduced that $T(n,k;m,m) = mn(H_{mn}-H_{mn-mk})$, which is achieved by any $[mn,mk]$ MDS code. On the other hand, if $k=1$ and $m>1$ then we achieve the random access version of the problem in~\cite{bar2023cover}, which was studied mainly for $a=1$. However, the value of $T(n,1;m,1)$ is still far from being solved. A lower bound states that for all $n$, $T(n,1;m,1)\geq n=\frac{n(n-m)}{m}(H_n-H_{n-m})$, while several code constructions verify that $T(n,1;m,1)< m$. For example, it was shown that there exists $n$ large enough such that $T(n,1;2,1)< 0.91\cdot 2$ and $T(n,1;3,1)< 0.89\cdot 3$ and if $m$ is a multiple of 4 then $T(n=2m,1;m,1)< 0.95\cdot m$. Based on the results~\cite{bar2023cover}, it is simple to deduce that $T(k,k;m,a) = mkH_{ak}$, and thus for the rest of the paper we assume that $n>k$. Several more results on this value and related problems have been studied lately in~\cite{GBRY24}, \cite{preuss2024sequencing}. 


\section{Random Access Expectation for a \\ Single File ($a=1$)}\label{sec:random}
This section studies the problem of optimizing the sample size for random access queries, 
where the user wishes to retrieve 1 file. Three coding schemes will be analyzed. 
\subsection{The Local MDS Scheme}\label{subsec:scheme1}
In this coding scheme, denoted as $\cC_1$, 
we employ an MDS code on each of the $m$ files separately and store each file in $n$ strands. Note that in this coding scheme, in order to decode any file, it is necessary and sufficient to retrieve any $k$ out of its $n$ encoded strands.

Our main goal in this section is to determine the expected number of samples for recovering any of the $m$ files while applying the coding scheme $\cC_1$. 
To analyze the performance of the coding scheme $\cC_1$, we let $t(h,i)$ be the random variable that denotes the number of samples required to progress from drawing $i$ to $i+1$ different strands out of the pool of $n$ encoded strands for the $h$-th file. Note that $t(h,i)$ follows a geometric distribution $t(h,i) \sim \text{Geo}\left(\frac{n-i}{mn}\right)$, where $\frac{n-i}{mn}$ is the probability of drawing the $(i+1)$-st strand, and consequently, the expected value, $\E\left[t(h,i)\right] = \frac{mn}{n-i}$. Furthermore, let $T(h,b)$ be the random variable representing the number of samples needed to progress from drawing $0$ to $b$ different strands of the $h$-th file. Hence, by definition $T(h,b) = \sum_{i=0}^{b-1} t(h,i)$ and thus 

\vspace{-2ex}
\begin{small}
   \begin{align}
\nonumber \E[T(h,b)] & = \E\left[\sum_{i=0}^{b-1} t(h,i)\right] = \sum_{i=0}^{b-1} \E\left[t(h,i)\right] & \\
& = \sum_{i=0}^{b-1} \frac{mn}{n-i} = mn(H_{n}-H_{n-b}). \label{eq:exp_C1}
\end{align} 
\end{small}


\begin{theorem}
    For any $1\leq k\leq n$ and $m\geq 1$, it holds that $$T(n,k;m,1)\leq \Tmax{\cC_1}{1} =mn (H_{n} - H_{n-k}).$$

\end{theorem}
\begin{proof}
    Assume without loss of generality that $F=\{1\}$. Note that by applying $b=k$ in (\ref{eq:exp_C1}) it holds that $\E[\nu_{(m,\{1\})}(\cC_1)] = \E[T(1,k)] = mn(H_{n} - H_{n-k})$. This also implies that  $$T(n,k;m,1)\leq\Tmax{\cC_1}{1} =mn  (H_{n} - H_{n-k}).\vspace{-3.5ex}$$
\end{proof}
\begin{corollary}
    For fixed $m$, $0 < R < 1$, $R=\frac{k}{n}$ for $n$ large enough it holds that $ \Tmax{\cC_1}{1}=mn \log\left(\frac{1}{1-R}\right)$.
Furthermore, for any fixed $m$ and $k$, it holds that, \vspace{-1ex}
$$\liminf_{n\to\infty}T(n,k;m,1) \leq \liminf_{n\to\infty}\Tmax{\cC_1}{1} = mk.\vspace{-1ex}$$
\end{corollary}

\begin{proof}
    Since $\liminf_{n\to\infty}\frac{H_n}{\log n}=1$, we can conclude that the following holds for $n$ large enough, 
\begin{align*}
 &\Tmax{\cC_1}{1} = mn(H_n -H_{n-k}) 
                \\ & = mn \left(\log(n) - \log(n-k) \right) 
                 \\&= mn \log\left( \frac{n}{n-k}\right)
                \\ & =  mn \log\left(\frac{1}{1-R}\right) 
\end{align*}
As for the limit,
\begin{align*}
    &\liminf_{n\to\infty}\Tmax{\cC_1}{1} = \liminf_{n\to\infty}mn(H_n-H_{n-k}) \\&=\liminf_{n\to\infty}mn\sum_{i=0}^{k-1}\frac{1}{n-i}=\liminf_{n\to\infty}m\sum_{i=0}^{k-1}\frac{n}{n-i}=mk.
\end{align*}   
\end{proof}

\subsection{The Global MDS Scheme}\label{subsec:scheme2}
In this coding scheme, denoted as $\cC_2$, we employ a systematic MDS code on the combined strands of the $m$ files. Hence, we store the $mk$  information strands into $mn$ encoded strands. In order to decode any of the $m$ files, it is necessary and sufficient to either retrieve all the systematic $k$ strands of the file, or any $mk$ out of $mn$ strands. The latter option decodes all $m$ files and in particular the required file.  

Our main goal is to find the expected number of samples to recover any of the $m$ files as defined in \autoref{prob:random:single} while applying the coding scheme $\cC_2$. Let $t(i)$ be the random variable that denotes the number of samples required to progress from drawing $i$ to $i+1$ different strands out of the pool of $mn$ encoded strands. Note that $t(i)$ follows a geometric distribution $t(i) \sim \text{Geo}\left(\frac{mn-i}{mn}\right)$, where $\frac{mn-i}{mn}$ is the probability of drawing the $(i+1)$-st strand, and thus $\E\left[t(i)\right] = \frac{mn}{mn-i}$. Furthermore, let $T(b)$ be the random variable representing the number of samples needed to progress from drawing $0$ to $b$ different strands. Hence, by definition $T(b) = \sum_{i=0}^{b-1} t(i)$ and

\vspace{-3ex}
    \begin{align}
\nonumber \E[T(b)] 
= \sum_{i=0}^{b-1} \E\left[t(i)\right] = \sum_{i=0}^{b-1} \frac{mn}{mn-i} = mn(H_{mn}-H_{mn-b}).\vspace{-1ex}
\end{align}



In order to analyze the collection process we will represent it as a discrete-time Markov chain. Let $X_b$ be a random variable that represents the state of the collection process \textbf{after} drawing $b$ different strands.
Indeed, the collection process satisfies the Markovian property, i.e., for a collection of $b$ states say $s_0, s_1, \ldots, s_{b-1}$ $\text{Pr}\left(X_{b}=s_{b} | X_0=s_0, \ldots, X_{b-1}=s_{b-1}\right) = \text{Pr}\left(X_{b}=s_{b} | X_{b-1}=s_{b-1}\right)$. For the setups under consideration, the states will be all compositions of different types of collected strands which will depend in general on the underlying coding scheme itself. We denote $\widehat{s}_i$ to be the sum of collected strands at state $s_i$.
Moreover, we let $M$ denote the transition matrix of this Markov chain where $M_{s_j, s_i} = \text{Pr}(X_{b}=s_i | X_{b-1}=s_{j} )$ for two states $s_i$ and $s_j$. Define $M_{s_j,s_i}^{(n)} = \text{Pr}(X_{n}=s_i | X_{0}=s_{j})$ 
which is the probability of collecting the additional strands from $s_j$ to the composition of collected strands in $s_i$ (i.e., $\widehat{s}_i-\widehat{s}_j=n$ strands)\footnote{At state $s_j$, we have two potential transitions: either remaining in the current state by drawing a previously collected strand or collect a new one and progressing to a new state. The variables $M$, and $M^{(n)}$ are analyzed and defined as the conditional probabilities of transitioning to a new state given new strands were drawn.}. Also, define the $n$-step transition probability matrix $M^{(n)}\triangleq(M_{s_j,s_i}^{(n)})$.
Note that $M_{s_j,s_i}^{(n)}=\sum_{s_y} M_{s_j,s_y}^{(n-1)} M_{s_y, s_i}$, where for $\widehat{s}_i-\widehat{s}_j\neq n$ then, $M_{s_j,s_i}^{(n)}=0$.
For shorthand, we refer to the initial state as $s_0$ (i.e., collect nothing from the pool of strands).

Our aim is to compute the expected hitting times for the absorbing states which will depend in general on the underlying coding scheme itself. This is established in the next theorem. 
\begin{theorem}\label{lem:random:single:C2 code a 1}
For any $1\leq k\leq n$ and $ m\geq 1$ it holds that $T(n,k;m,1)\leq\Tmax{\cC_2}{1}$ and
\vspace{-1.5ex}
    \begin{small}
    \begin{align*}
        \Tmax{\cC_2}{1}=&mn\left(H_{mn}\hspace{-0.4ex}-\hspace{-0.4ex}H_{mn-mk}\right)\\
        &\hspace{-4.5ex}-\hspace{-0.5ex} \frac{mn}{\binom{mn}{k}}\hspace{-0.5ex}\Bigg(\hspace{-0.5ex}\sum_{j=0}^{mk-k-1}\hspace{-1ex} \binom{k\hspace{-0.5ex}-\hspace{-0.5ex}1\hspace{-0.5ex}+\hspace{-0.5ex}j}{k\hspace{-0.5ex}-\hspace{-0.5ex}1} H_{mn-(k+j)} \hspace{-0.5ex}-\hspace{-0.5ex} H_{mn-mk}\binom{mk\hspace{-0.5ex}-\hspace{-0.5ex}1}{k}\hspace{-0.5ex}\Bigg).
    \end{align*}
\end{small}
\end{theorem}
\begin{proof}

Assume without loss of generality that $F=\{1\}$.
\begin{itemize}
    \item \textbf{States definition:} The set of states is $\cS_1=\set{(i,j)}{0\leq i\leq k, 0\leq j\leq mn-k}$, where $i$ is the number of strands drawn from the $k$ systematic strands of the first file, and $j$ is the number of strands that were drawn from the other $mn-k$ strands. 
    \item \textbf{Transition matrix:} 

    The valid transitions in $M$ (i.e., $M_{(i_1,j_1),(i_2,j_2)}\neq 0$) and their values are

    \vspace{-2ex}
    \begin{small}
        \begin{align*}
        &M_{(i,j),(i+1,j)} = \frac{k-i}{mn-(i+j)}, M_{(i,j), (i,j+1)} = \frac{mn-k-j}{mn-(i+j)}.
    \end{align*}
    \end{small}
    The next claim provides a closed formula for $M_{(0,0),(i,j)}^{(i+j)}$ which holds for the non-absorbing states. 
    \begin{claim} 
    \label{claim:prg:C2 a 1}
  
         $M_{(0,0),(i,j)}^{(i+j)}=\frac{\binom{i+j}{i} \binom{mn-(i+j)}{k-i}}{\binom{mn}{k}}$.
    \end{claim}
    \begin{proof}
    At state $(i,j)$, we have $\binom{k}{i}$ options to choose $i$ systematic strands and $\binom{mn-k}{j}$ options to choose $j$ strands from the rest of the pool, considering all possibilities for drawing a total of $(i+j)$ strands out of the $mn$ strands, which is $\binom{mn}{i+j}$.

        \begin{align*}
        M_{(0,0),(i,j)}^{(i+j)} &=\frac{\binom{k}{i} \binom{mn-k}{j}}{\binom{mn}{i+j}} = \frac{\frac{(k)! (mn-k)!}{(k-i)! i! (mn-k-j)! j!}}{\frac{(mn)!}{(i+j)! (mn-(i+j))!}} \\&=\frac{\binom{i+j}{i} \binom{mn-(i+j)}{k-i}}{\binom{mn}{k}}
    \end{align*}
    \end{proof}
    \vspace{-2ex}
    \item \textbf{Absorbing states:}
    These are the states that allow the recovery of the first file, so the drawing process ends. In coding scheme $\cC_2$ the absorbing states are those where we either drew the $k$ systematic strands of the first file or any $mk$ different strands from the pool of $mn$ strands. We denote $\Theta_1, \Theta_2$ as the set of absorbing states corresponding to the first, second option, respectively. That is,
    \begin{align*}
        &\Theta_1\triangleq\set{(k,j)}{0\leq j\leq mk-k-1},\\ &\Theta_2\triangleq\set{(i,mk-i)}{0\leq i\leq k}.
    \end{align*}
    Our approach involves investigating the transient states prior to absorption since those states determine a specific absorbing state. Denote $\cG_1, \cG_2$ the set of states reachable from a non-absorbing state to an absorbing one corresponding to $\Theta_1, \Theta_2$.
    \begin{align*}
        &\cG_1\triangleq\set{(k-1,j)}{0\leq j\leq mk-k-1},\\ &\cG_2\triangleq\set{(i,mk-i-1)}{0\leq i\leq k-1}.
    \end{align*}
    
    Note that:
    \begin{enumerate}
        \item Given $g=(k-1,j)\in \cG_1$  there exists 1 corresponding absorbing state $\theta = (k,j) \in \Theta_1$, (since $(k,j-1)$ is also an absorbing state), thus the probability of reaching the absorbing state $\theta$ is, 
        $$M_{(0,0),g}^{(\widehat{g})}\cdot M_{g,\theta}.$$
    \item Given $g=(i,mk-i-1)\in \cG_1$  there exists 2 corresponding absorbing states $\theta_1 = (i,mk-i), \theta_2 =  (i+1,mk-i-1) \in \Theta_2$, thus the probability of reaching the absorbing states $\theta_1, \theta_2$ is, 
    \begin{align*}
        M_{(0,0),g}^{(\widehat{g})}\cdot (M_{g,\theta_1} +M_{g,\theta_2})=M_{(0,0),g}^{(\widehat{g})}.
    \end{align*}

    \end{enumerate}
    \item \textbf{The expectation:}
    In order to calculate $\E[\nu_{(m,\{1\})}(\cC_2)]$, we let $Y$ be the random variable representing in which absorbing state the collection process ends. The expectation $\E[\nu_{(m,\{1\})}(\cC_2)]$ is conditioned on $Y$. Hence,
    \begin{small}
        \begin{align}
        \nonumber&\E[\nu_{(m,\{1\})}(\cC_2)]=\E_{Y}[\E[\nu_{(m,\{1\})}(\cC_2)|Y]]\\&\nonumber=\sum_{\theta\in \Theta_1\cup \Theta_2 } \text{Pr}(Y=\theta)\cdot \E[\nu_{(m,\{1\})}(\cC_2)|Y=\theta]\\&
        \nonumber= \sum_{\theta\in \Theta_1 } \text{Pr}(Y=\theta)\cdot \E[\nu_{(m,\{1\})}(\cC_2)|Y=\theta]\\&
        \nonumber +\sum_{\theta\in \Theta_2 } \text{Pr}(Y=\theta)\cdot \E[\nu_{(m,\{1\})}(\cC_2)|Y=\theta]\\&
        \nonumber=^{(*)}\sum_{j=0}^{mk-k-1} M_{(0,0),(k,j)}^{(k+j)}\cdot M_{(k-1,j),(k,j)}\E[T(k+j)]\\&\nonumber+\sum_{i=0}^{k-1} M_{(0,0),(i,mk-i-1)}^{(mk-1)}\E[T(mk)]\\
        \nonumber &=mn\left(H_{mn}\hspace{-0.4ex}-\hspace{-0.4ex}H_{mn-mk}\right)\\&\nonumber-\frac{mn}{\binom{mn}{k}}\Bigg(\sum_{j=0}^{mk-k-1} \binom{k\hspace{-0.3ex}-\hspace{-0.3ex}1+j}{k\hspace{-0.3ex}-\hspace{-0.3ex}1} H_{mn-(k+j)}\\
         &\hspace{4ex} - H_{mn-mk} \cdot \binom{mk-1}{k}\Bigg),\label{eq:excpc2ais1}
    \end{align}
    \end{small}
     where $(*)$ follows from the definitions of $\cG_1, \cG_2$ thus we iterate over all absorbing states 
     \begin{align*}
         &\sum_{\theta\in \Theta_1}\text{Pr}(Y=\theta) \hspace{-0.5ex}=\hspace{-2ex}\sum_{g=(k-1,j)\in \cG_1} \hspace{-2.5ex}M_{(0,0),(k-1,j)}^{(k-1+j)}\cdot M_{(k-1,j),(k,j)},\\&
         \sum_{\theta\in \Theta_2}\text{Pr}(Y=\theta) \hspace{-0.5ex}= \hspace{-2ex}\sum_{g=(i,mk-i-1)\in \cG_1}\hspace{-2.5ex} M_{(0,0),(i,mk-i-1)}^{(mk-1)}.
     \end{align*}

    The full proof of (\ref{eq:excpc2ais1}) can be found in Appendix \ref{Appendix_expectence_c2} by applying $a=1$.
    Since the code is symmetric it implies that
    \vspace{-1ex}
    \begin{small}
        \begin{align*}
            \E[\nu_{(m,\{1\})}(\cC_2)]=\Tmax{\cC_2}{1}.
        \end{align*}
    \end{small}
    \end{itemize}
\end{proof}

\begin{corollary}
For any fixed $m$ and $k\geq2$, it holds that 
$$\liminf_{n\to\infty}T(n,k;m,1)\leq\liminf_{n\to\infty}\Tmax{\cC_2}{1} = mk.$$
\end{corollary}
\begin{proof}
\begin{small}
    \begin{align*}
    &\liminf_{n\to\infty}\Tmax{\cC_2}{1} \\&=\liminf_{n\to\infty}mn\left(H_{mn}\hspace{-0.4ex}-\hspace{-0.4ex}H_{mn-mk}\right)\\&-\liminf_{n\to\infty} \frac{mn}{\binom{mn}{k}}\Bigg(\sum_{j=0}^{mk-k-1} \binom{k\hspace{-0.3ex}-\hspace{-0.3ex}1+j}{k\hspace{-0.3ex}-\hspace{-0.3ex}1} H_{mn-(k+j)}\\
         &\hspace{4ex} - H_{mn-mk} \cdot \binom{mk-1}{k}\Bigg).
\end{align*}
\end{small}
We will show that the first limit is $mk$ and the second is 0.
As for the first part, we have that: 
\vspace{-2ex}
\begin{small}
    \begin{align*}
    &\liminf_{n\to\infty}mn\left(H_{mn}\hspace{-0.4ex}-\hspace{-0.4ex}H_{mn-mk}\right)=\liminf_{n\to\infty}mn\sum_{i=0}^{mk-1}\frac{1}{mn-i}\\&
        =\liminf_{n\to\infty}\sum_{i=0}^{mk-1}\frac{mn}{mn-i}=mk.
    \end{align*}
\end{small}
As for the second part, it is known that $\liminf_{n\to\infty}\frac{H_n}{\log n}=1$. For fixed $m,k$ we can deduce that the numerator behaves as $\mathcal{O}(n\log{}n)$ and the denominator as ${\mathcal{O}(n^k)}$. Then, there exist some constants $C_1,C_2$ (which depend on $m$ and $k$) such that the following holds,

\vspace{-2ex}
\begin{small}
    \begin{align*}
        &\liminf_{n\to\infty}\frac{mn}{\binom{mn}{k}}\Bigg(\sum_{j=0}^{mk-k-1} \binom{k\hspace{-0.3ex}-\hspace{-0.3ex}1+j}{k\hspace{-0.3ex}-\hspace{-0.3ex}1} H_{mn-(k+j)}\\
         &\hspace{4ex} - H_{mn-mk} \cdot \binom{mk-1}{k}\Bigg)\\&
         =\liminf_{n\to\infty} \frac{C_2\log n}{C_1n^{k-1}} = 0.
    \end{align*}
\end{small}
Hence,
\vspace{-2ex}
\begin{small}
    \begin{align*}
        &\liminf_{n\to\infty}\Tmax{\cC_2}{1}=mk+0=mk.
    \end{align*}
\end{small}
\vspace{-2ex}
\end{proof}

\subsection{The Partial MDS Scheme}\label{subsec:scheme3}
In this section, we consider the case where the underlying retrieval code, denoted $\cC_3$, is a partial-MDS (PMDS) code and we apply to $m=2$ files. We briefly review the definition of a $[r;s]$ code before proceeding.

\begin{definition}\label{def:PMDS} Let $C$ be a linear $[m_P n_P,m_P(n_P-r_P)-2s]$ code over a field such that if codewords are taken row-wise as $m_P \times n_P$ arrays, each row belongs to an $[n_P,n_P-r_P,r_P+1]$ MDS code. Given $\sigma_1, \ldots, \sigma_t$ such that for $j\in [t]$, $\sigma_j \geq 1$, we say that $C$ is an $(r_P;\sigma_1,\ldots,\sigma_t)$-erasure correcting code if, for any $1\leq i_1 < \cdots < i_t \leq m$, $C$ can correct up to $\sigma_j+r_P$ erasures in row $i_j$ of an array in $C$. We say that $C$ is an $(r_P;2s)$ PMDS code if, for every $(\sigma_1,\ldots,\sigma_t)$ where $\sum_{j=1}^t \sigma_j=2s$, $C$ is an $(r_P;\sigma_1,\ldots,\sigma_t)$-erasure correcting code.
\end{definition}
\vspace{-1ex}
Constructions of $(r;s)$ PMDS codes have been shown to exist for all $r$ and $s$ provided large enough field sizes \cite{gabrys2018constructions}.
For the purposes of our problem, we assume that $m_P=2$ and that the information dimension of each of the two files we encode is $k=\frac{1}{2} \left( 2(n_P - r_P) - 2s \right) = n_P - r_P - s$ and where $n=n_P$ so that we can interpret for instance the information for the first file being contained in a systematic code that appears in the first row and the information for the second file appearing as the second row in our codeword according to the previous definition. Then according to Definition~\ref{def:PMDS}, we can recover file $1$ in the following ways:
\begin{enumerate}
\item File $1$ can be recovered by collecting the $k$ systematic strands for File $1$ which appear in the first row.
\item File $1$ can be recovered by collecting $n_P - r_P = k + s$ strands out of the $n$ strands in the first row.
\item File $1$ can be recovered by collecting $2k$ distinct strands whereby at least $n_P-r_P-2s = k-s$ and at most $n_P-r_P=k+s$ originate from the first row.
\end{enumerate}
Since the code is symmetric, the ways for recovering file 2 mirror those for recovering file 1.
By using the same notations of $t(i), T(b)$ and the Markov chain properties as mentioned in \autoref{subsec:scheme2}, the next theorem is proved similarly.

\begin{theorem}
\label{lem:random:single:C3 code}
For any $1\leq k\leq n$ and $ m=2$, $T(n,k;2,1)\leq\Tmax{\cC_3}{1}$ and:

\vspace{-2ex}
\begin{scriptsize}
\begin{align*}
    &\Tmax{\cC_3}{1}=2n\sum_{j=0}^{s}\sum_{h=0}^{k-j}\frac{k \binom{n-k}{j} \binom{n}{h}}{\binom{2n}{k-1+j+h}}\cdot\frac{H_{2n}-H_{2n-(k+j+h)}}{2n-(k-1+j+h)}\\
    &\hspace{-0.5ex}+\hspace{-0.5ex}
    2n\sum_{i=1}^{k-1}\sum_{h=0}^{k-s} \frac{\binom{k}{i-1} \binom{n-k}{k+s-i} \binom{n}{h}}{\binom{2n}{k+s-1+h}} \cdot\frac{(k-i+1)(H_{2n}\hspace{-0.5ex}-\hspace{-0.5ex}H_{2n-(k+s+h)})}{2n-(k+s-1+h)}\\
    &\hspace{-0.5ex}+\hspace{-0.5ex}2n\sum_{i=0}^{k-1}\sum_{h=0}^{k-s}\frac{\binom{k}{i} \binom{n-k}{k+s-i-1} \binom{n}{h}}{\binom{2n}{k+s-1+h}}\cdot\frac{(n\hspace{-0.5ex}-\hspace{-0.5ex}2k\hspace{-0.5ex}-\hspace{-0.5ex}s\hspace{-0.5ex}+\hspace{-0.5ex}i\hspace{-0.5ex}+\hspace{-0.5ex}1)(H_{2n}\hspace{-0.5ex}-\hspace{-0.5ex}H_{2n-(k+s+h)})}{2n-(k+s-1+h)}\\
    &\hspace{-0.5ex}+\hspace{-0.5ex}2n\sum_{i=1}^{k-s}\sum_{h=k+s+1}^{n} \frac{\binom{k}{i-1} \binom{n-k}{k-s-i} \binom{n}{h}}{\binom{2n}{k-s-1+h}} \cdot\frac{(k\hspace{-0.5ex}-\hspace{-0.5ex}i\hspace{-0.5ex}+\hspace{-0.5ex}1))(H_{2n}\hspace{-0.5ex}-\hspace{-0.5ex}H_{2n-(k-s+h)})}{2n-(k-s-1+h)}\\
    &\hspace{-0.5ex}+\hspace{-0.5ex}2n\sum_{i=0}^{k-s}\sum_{h=k+s+1}^{n} \hspace{-2.5ex}\frac{\binom{k}{i} \binom{n-k}{k-s-i-1} \binom{n}{h}}{\binom{2n}{k-s-1+h}} \hspace{-0.5ex}\cdot\hspace{-0.5ex}\frac{(n\hspace{-0.5ex}-\hspace{-0.5ex}2k\hspace{-0.5ex}+\hspace{-0.5ex}s\hspace{-0.5ex}+\hspace{-0.5ex}i\hspace{-0.5ex}+\hspace{-0.5ex}1)(H_{2n}\hspace{-0.5ex}-\hspace{-0.5ex}H_{2n-(k-s+h)})}{2n-(k-s-1+h)}\\
    &\hspace{-0.5ex}+\hspace{-0.5ex}2n\sum_{i=0}^{k-1}\sum_{j=\max(0,k-s-i)}^{k+s-i-1}\frac{\binom{k}{i} \binom{n-k}{j} \binom{n}{2k-(i+j)}}{\binom{2n}{2k}}(H_{2n}-H_{2n-2k}).
\end{align*}  
\end{scriptsize}
\end{theorem} 
\begin{proof}
Assume without loss of generality that $F=\{1\}$.
\begin{itemize}
    \item \textbf{States definition:} The set of states is $\cS_2=\set{(i,j,h)}{0\leq i\leq k, 0\leq j\leq n-k, 0\leq h\leq n}$, where $i$ is the number of strands drawn from the $k$ systematic strands of the first file, $j$ is the number of strands drawn from the $n-k$ encoded non-systematic strands of the first file, and $h$ is the number of strands that were drawn from the other file.
    \item \textbf{Transition matrix:} 
    The valid transitions in $M$ (i.e., $M_{(i_1,j_1,h_1),(i_2,j_2,h_2)}\neq 0$) and their values are:
    \begin{align*}
        &M_{(i,j,h),(i+1,j,h)} = \frac{k-i}{2n-(i+j+h)}\triangleq M_1,\\
        &M_{(i,j,h),(i,j+1,h)} = \frac{n-k-j}{2n-(i+j+h)}\triangleq M_2,\\
        &M_{(i,j,h),(i,j,h+1)} = \frac{n-h}{2n-(i+j+h)}\triangleq M_3.
    \end{align*}

    The next claim provides a closed formula for $M_{(i,j,h)}^{(i+j+h)}$ which holds for the non-absorbing states.
    \begin{claim} 
    \label{claim:prg:C3 m2 a 1}
         $M_{(0,0,0),(i,j,h)}^{(i+j+h)}=\frac{\binom{k}{i} \binom{n-k}{j} \binom{n}{h}}{\binom{2n}{i+j+h}}$     
    \end{claim}
    \begin{proof}

    At the state $(i,j,h)$, we have $\binom{k}{i}$ options to choose $i$ systematic strands, $\binom{n-k}{j}$ options to choose $j$ strands from the other non-systematic encoded strands of the first file, and $\binom{n}{h}$  options to choose $h$ strands from the second file. Considering all possibilities for drawing a total of $(i+j+h)$ strands out of the $mn$ strands, which is $\binom{2n}{i+j+h}$.
    \end{proof}
    \item \textbf{Absorbing states:} In scheme code $\cC_3$ the absorbing states are divided into 4 categories: end with the systematic $k$ strands, end with any $k+s$ strands of the first file, end with more than $2k$ strands of both files and, end with exactly $2k$ strands of both files. Denote as:

   \begin{small}
       \begin{align*}
        &\Theta_1\triangleq\set{(k,j,h)}{0 \leq j \leq s, 0 \leq h \leq k-j}\\ &\Theta_2\triangleq\set{(i,k+s-i,h)}{1 \leq i \leq k-1, 0 \leq h \leq k-s }\\
        &\Theta_3\triangleq\set{(i,k+s-i,h)}{0 \leq i \leq k-1, 0 \leq h \leq k-s }\\
        &\Theta_4\triangleq\set{(i,k-s-i,h)}{1 \leq i \leq k-s, k+s < h \leq n }\\
        &\Theta_5\triangleq\set{(i,k-s-i,h)}{0 \leq i \leq k-s, k+s < h \leq n }\\
        &\Theta_6\triangleq\{(i,j,2k-(i+j))|\\&0 \leq i \leq k-1, \max\{0, k-s-i\} \leq j \leq k+s-i-1\}.
    \end{align*}
   \end{small}

    Note that:
    \begin{enumerate}
        \item Given \begin{small}
            $\theta\in \Theta_1$ $M_{(0,0,0),(\theta)}^{(\widehat{\theta)}}=M_{(0,0,0),(k-1,j,h)}^{(k-1+j+h)}\cdot M_1$.
        \end{small}
        \item Given \begin{small}
            $\theta\in \Theta_2$ $M_{(0,0,0),(\theta)}^{(\widehat{\theta)}}=M_{(0,0,0),(i-1,k+s-i,h)}^{(k+s-1+h)}\cdot M_1$.
        \end{small}
        \item Given \begin{small}
            $\theta\in \Theta_3$ $M_{(0,0,0),(\theta)}^{(\widehat{\theta)}}=M_{(0,0,0),(i, k+s-i-1, h)}^{(k+s-1+h)}\cdot M_2$.
        \end{small} 
        \item Given \begin{small}
            $\theta\in \Theta_4$  $M_{(0,0,0),(\theta)}^{(\widehat{\theta)}}=M_{(0,0,0),(i-1, k-s-i, h)}^{(k-s-1+h)}\cdot M_1$.
        \end{small}
        \item Given \begin{small}
            $\theta\in \Theta_5$  $M_{(0,0,0),(\theta)}^{(\widehat{\theta)}}=M_{(0,0,0),(i, k-s-i-1, h)}^{(k-s-1+h)}\cdot M_2$.
        \end{small}
    \end{enumerate}

    \item \textbf{The expectation:}
    In order to calculate $\E[\nu_{(2,\{1\})}(\cC_3)]$, we let $Y$ be the random variable representing in which absorbing state the collection process ends. The expectation $\E[\nu_{(m,\{1\})}(\cC_2=3)]$ is conditioned on $Y$. Hence,

    \begin{scriptsize}

            \begin{align*}
        &\E[\nu_{(2,\{1\})}(\cC_3)]=\E_{Y}[\E[\nu_{(2,\{1\})}(\cC_3)|Y]]
        \\&=\sum_{\theta\in \cup_{i=1}^{6} \Theta_i } M_{(0,0,0),(\theta)}^{(\widehat{\theta})}\cdot\E[\nu_{(2,\{1\})}(\cC_3)|Y=\theta]
        \\&= \sum_{\theta\in \cup_{i=1}^{6} \Theta_i } M_{(0,0,0),(\theta)}^{(\widehat{\theta})}\E[T(\widehat{\theta})]
        \\&= \sum_{\theta\in \Theta_1} M_{(0,0,0),(k-1, j, h)}^{(k-1+j+h)}\cdot M_1\cdot \E[T(k+j+h)]
        \\&+
        \sum_{\theta\in \Theta_2 } M_{(0,0,0),(i, k+s-i-1, h)}^{(k+s-1+h)}\cdot M_2\cdot \E[T(k+s+h)]
        \\&+\sum_{\theta\in \Theta_3 }M_{(0,0,0),(i, k+s-i-1, h)}^{(k+s-1+h)}\cdot M_2\cdot \E[T(k+s+h)]
        \\&+
        \sum_{\theta\in \Theta_4 }M_{(0,0,0),(i-1, k-s-i, h)}^{(k-s-1+h)} \cdot M_1\cdot \E[(k-s+h)]
        \\&+\sum_{\theta\in \Theta_5 }M_{(0,0,0),(i, k-s-i-1, h)}^{(k-s-1+h)} \cdot M_2\cdot \E[T(k-s+h)]
        \\&+\sum_{\theta\in \Theta_6 } M_{(0,0,0),(i,j,2k-(i+j))}^{(2k)}\cdot \E[T(2k)]
        \\&=2n\sum_{j=0}^{s}\sum_{h=0}^{k-j}\frac{k \binom{n-k}{j} \binom{n}{h}}{\binom{2n}{k-1+j+h}}\cdot\frac{H_{2n}-H_{2n-(k+j+h)}}{2n-(k-1+j+h)}
        \\&\hspace{-0.5ex}+\hspace{-0.5ex}
    2n\sum_{i=1}^{k-1}\sum_{h=0}^{k-s} \frac{\binom{k}{i-1} \binom{n-k}{k+s-i} \binom{n}{h}}{\binom{2n}{k+s-1+h}} \cdot\frac{(k-i+1)(H_{2n}\hspace{-0.5ex}-\hspace{-0.5ex}H_{2n-(k+s+h)})}{2n-(k+s-1+h)}
    \\&\hspace{-0.5ex}+\hspace{-0.5ex}2n\sum_{i=0}^{k-1}\sum_{h=0}^{k-s}\frac{\binom{k}{i} \binom{n-k}{k+s-i-1} \binom{n}{h}}{\binom{2n}{k+s-1+h}}\cdot\frac{(n\hspace{-0.5ex}-\hspace{-0.5ex}2k\hspace{-0.5ex}-\hspace{-0.5ex}s\hspace{-0.5ex}+\hspace{-0.5ex}i\hspace{-0.5ex}+\hspace{-0.5ex}1)(H_{2n}\hspace{-0.5ex}-\hspace{-0.5ex}H_{2n-(k+s+h)})}{2n-(k+s-1+h)}
    \\&\hspace{-0.5ex}+\hspace{-0.5ex}2n\sum_{i=1}^{k-s}\sum_{h=k+s+1}^{n} \frac{\binom{k}{i-1} \binom{n-k}{k-s-i} \binom{n}{h}}{\binom{2n}{k-s-1+h}} \cdot\frac{(k\hspace{-0.5ex}-\hspace{-0.5ex}i\hspace{-0.5ex}+\hspace{-0.5ex}1))(H_{2n}\hspace{-0.5ex}-\hspace{-0.5ex}H_{2n-(k-s+h)})}{2n-(k-s-1+h)}
    \\&\hspace{-0.5ex}+\hspace{-0.5ex}2n\sum_{i=0}^{k-s}\sum_{h=k+s+1}^{n} \hspace{-2.5ex}\frac{\binom{k}{i} \binom{n-k}{k-s-i-1} \binom{n}{h}}{\binom{2n}{k-s-1+h}} \hspace{-0.5ex}\cdot\hspace{-0.5ex}\frac{(n\hspace{-0.5ex}-\hspace{-0.5ex}2k\hspace{-0.5ex}+\hspace{-0.5ex}s\hspace{-0.5ex}+\hspace{-0.5ex}i\hspace{-0.5ex}+\hspace{-0.5ex}1)(H_{2n}\hspace{-0.5ex}-\hspace{-0.5ex}H_{2n-(k-s+h)})}{2n-(k-s-1+h)}
    \\&\hspace{-0.5ex}+\hspace{-0.5ex}2n\sum_{i=0}^{k-1}\sum_{j=\max(0,k-s-i)}^{k+s-i-1}\frac{\binom{k}{i} \binom{n-k}{j} \binom{n}{2k-(i+j)}}{\binom{2n}{2k}}(H_{2n}-H_{2n-2k}).
    \end{align*}
    \end{scriptsize}

    Since the code is symmetric it implies that
    $$\E[\nu_{(2,\{1\})}(\cC_3)]=\Tmax{\cC_3}{1}.$$
    
\end{itemize}
\end{proof}

\section{Lower Bounds}\label{sec:lowerbound}
In this section, we present two lower bounds on the value of $T(n,k;m,1)$. The first bound does not depend on $n$, while the second presents an improvement considering it.
\begin{lemma} \label{lm:lowerboundmnmk}  
For any $n,k,m$ it holds $T(n,k;m,1) \geq \frac{k(m+1)}{2}$.
\end{lemma}

\begin{proof}
Recall that $\cU, \cX$ represent the information and encoded strands of the $m$ files respectively. Consider the scenario where all terms in $\cU$ have been encoded into the codeword $\cX$. Represent every sequence of reads as a vector $\bfv\in[mn]^*$, and for each $\bfv$, denote $n_i(\bfv)$ for $i\in[m]$ as the minimum read index $h$ that facilitates the retrieval of the $i$-th file $U_i$. With each new sample obtained in the sequence of reading the strands, the recovery of at most one new information strand is possible. Consequently, the minimum number of strands required to recover each file is $k$. In the best-case scenario, assuming that the initial $k$ strands pertain to one file, the subsequent $k$ strands relate to another file among the $m$ files, and so forth, assuming some permutation on the files $\sigma$ we obtain:
\begin{align*}
    &\sum_{i=1}^m n_{\sigma(i)}(\bfv)= n_1(\bfv)+n_2(\bfv)+\cdots +n_m(\bfv)\\&\geq\sum_{i=1}^m ki = \frac{km(m+1)}{2}.
\end{align*}

Hence, it follows that, $\sum_{i=1}^m\nu_{(m,\{i\})}(\cC)\geq \frac{km(m+1)}{2}$ and therefore 
$$\E\left[\sum_{i=1}^m \nu_{(m,\{i\})}(\cC)\right]=  \sum_{i=1}^m\E[ \nu_{(m,\{i\})}(\cC)] \geq \frac{km(m+1)}{2}.$$ In particular, there exists $i\in[m]$ for which $\E[ \nu_{(m,\{i\})}(\cC)] \geq \frac{k(m+1)}{2}$, i.e., $\Tmax{\cC}{1}\geq T(n,k;m,1) \geq \frac{k(m+1)}{2}$.     
\end{proof}

In order to consider the effect of $n$ on the value of $T(n,k;m,1)$, we obtain a tighter lower bound on $T(n,k;m,1)$ compared with \autoref{lm:lowerboundmnmk}. 
\begin{theorem}\label{th: ramdom access: better lower bound}
   For any $n,k,m$ it holds that
   \vspace{-1.3ex}
    \[
    T(n,k;m,1) \ge mnH_{mn}-n\sum_{i=1}^{m}H_{mn-ki}\geq \frac{k(m+1)}{2}.
    \]
\end{theorem}
\begin{proof}
Let us use the same notations
as in the proof of \autoref{lm:lowerboundmnmk}. Additionally, define $t_j(\cC)$ to be the random variable that governs the time to collect the $j$-th new sample (after collecting the previous one). Clearly, we have that
\[
\sum_{i=1}^m \nu_{(m,\{i\})}(\cC) = \sum_{i=1}^m n_i(v)\ge \sum_{i=1}^m \sum_{j=1}^{ki} t_j(v).
\]

Hence, 
    \begin{align*}
        &\sum_{i=1}^m \E \left[\nu_{(m,\{i\})}(\cC)\right] = \E\left[\sum_{i=1}^m \nu_{(m,\{i\})}(\cC)\right]  \\&\ge\E\left[\sum_{i=1}^m \sum_{j=1}^{ki} t_j(\cC)\right] = \sum_{i=1}^m\sum_{j=1}^{ki} \E\left[t_j(\cC)\right].
    \end{align*}

    Note that for any $j\in[mk]$, we have that $t_j(\cC)$ is a geometric random variable with success probability $p_j=\frac{mn-(j-1)}{mn}$  and so $\E[t_j(\cC)] = \frac{mn}{mn-(j-1)}$, and
    \begin{align*}
    &\sum_{i=1}^m \E \left[\nu_{(m,\{i\})}\right]  \ge  \sum_{i=1}^m \sum_{j=1}^{ki} \E\left[t_j(\cC)\right] 
    = \sum_{i=1}^m \sum_{j=1}^{ki} \frac{mn}{mn-(j-1)}\\
    & = mn \sum_{i=1}^m \left(\frac{1}{mn-ki+1} + \frac{1}{mn-ki+2} + \ldots + \frac{1}{mn}\right) \\
    &= mn \sum_{i=0}^{m-1} (m-i) \left( \sum_{j=0}^{k-1}\frac{1}{mn-ki -j}\right)\\
    &=mn \sum_{i=0}^{m-1} (m-i)(H_{mn-ki}-H_{mn-k(i+1)})\\
    &= m^2n(H_{mn}-H_{mn-mk})\\&-mn \sum_{i=0}^{m-1}i(H_{mn-ki}-H_{mn-k(i+1)})\\
    &= m^2n(H_{mn}-H_{mn-mk})\\&-mn \sum_{i=1}^{m-1}H_{mn-ki}+ mn(m-1)H_{mn-mk}\\
    &= m^2nH_{mn}-mn\sum_{i=1}^{m}H_{mn-ki}.
    \end{align*}
    Hence we have that 
    \begin{align*}
    &\frac{1}{m}  \sum_{i=1}^m \E \left[\nu_{(m,\{i\})}(\cC)\right] \geq mnH_{mn}-n\sum_{i=1}^{m}H_{mn-ki}.
    \end{align*}
    In particular, there exists $i\in[m]$ for which $\E[\nu_{(m,\{i\})}(\cC)]  \geq  mnH_{mn}-n\sum_{i=1}^{m}H_{mn-ki}$, i.e., $\Tmax{\cC}{1}\geq T(n,k;m,1)\geq mnH_{mn}-n\sum_{i=1}^{m}H_{mn-ki}.$ 

Next we will proof that $mnH_{mn}-n\sum_{i=1}^{m}H_{mn-ki}\geq \frac{k(m+1)}{2}$.
\begin{small}
  \begin{align*}
     &mnH_{mn}-n\sum_{i=1}^{m}H_{mn-ki}
     \\&=n(H_{mn}-H_{mn-k}+H_{mn}-H_{mn-2k}+\dots+H_{mn}-H_{mn-mk})\\&
     =n(\sum_{i=0}^{k-1}\frac{1}{mn-i}+\sum_{i=0}^{2k-1}\frac{1}{mn-i}+\dots+\sum_{i=0}^{mk-1}\frac{1}{mn-i})\\&
     =\sum_{j=1}^{m}\sum_{i=0}^{kj-1}\frac{n}{mn-i}\geq\sum_{j=1}^{m}\frac{1}{m}\cdot kj\\&= \frac{km(1+m)}{2m} = \frac{k(m+1)}{2}.
\end{align*}   
\end{small}
\end{proof}
The asymptotic behavior of this bound is given in the next corollary.
\begin{corollary}
    For fixed $m, k, R=\frac{k}{n}$ it holds that $\lim_{n\rightarrow\infty}T(n,k;m,1) \ge\lim_{n\rightarrow\infty}(mnH_{mn}-n\sum_{i=1}^{m}H_{mn-ki}) = \frac{k(m+1)}{2}$.
        Also, for fixed $0<R<1$, $\lim_{n\rightarrow\infty}\frac{T(n,k;m,1)}{n}\geq \lim_{n\rightarrow\infty}mH_{mn}-\sum_{i=1}^{m}H_{mn-ki} = \frac{R(m+1)}{2}$.
\end{corollary}
\begin{proof}
    The proofs follow directly from \autoref{th: ramdom access: better lower bound}.
\end{proof}

\section{Comparisons and Evaluations}\label{sec:compare}

In this section, we will conduct a comparative analysis of the coding schemes introduced in \autoref{sec:random}, focusing on their expected retrieval time, variance, and probability distribution. Our evaluation will commence with a comparison of expected retrieval times. Then, we will present and discuss the simulation results of the three coding schemes. Finally, we will conclude which of the schemes is superior.
This knowledge proves pivotal for the optimization of DNA storage systems, as our objective is to minimize the number of samples required for file recovery. 

\begin{figure}[ht]
        \includegraphics[width=1\linewidth, height=2in]{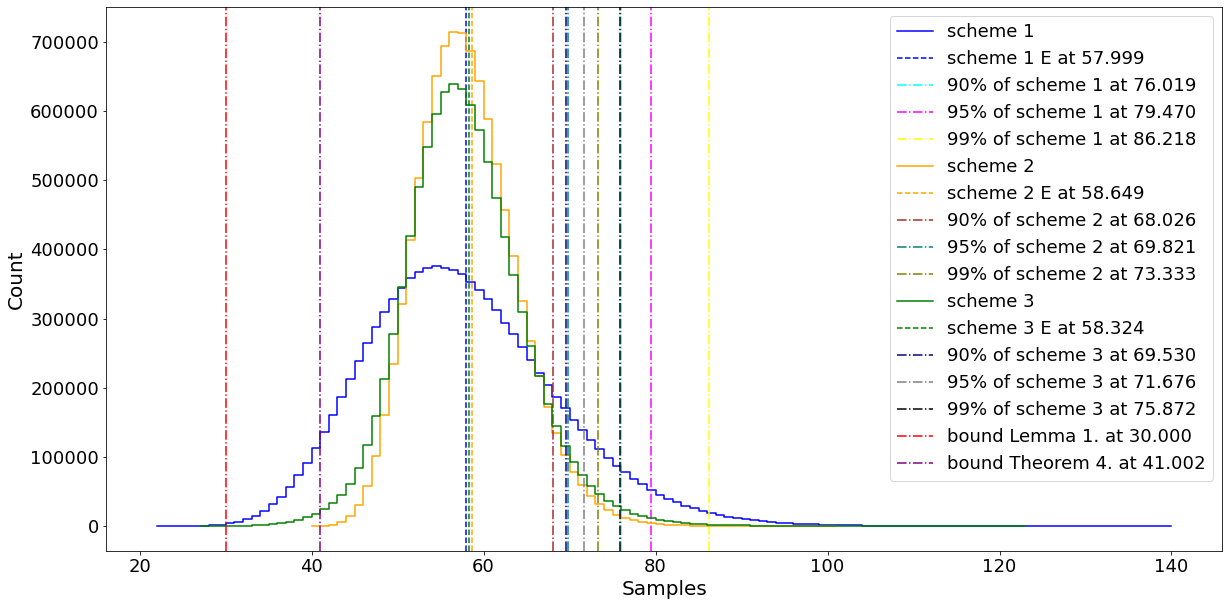}
        \caption{Illustrates the distribution of necessary sample sizes for file recovery and confidence levels of 90\%, 95\%, and 99\% across 3 coding schemes. And the lower bounds specified in \autoref{th: ramdom access: better lower bound} and in \autoref{lm:lowerboundmnmk}.}
        \label{fig:sim}
\end{figure}

First, we note that for two files ($m=2$) the first coding scheme is superior of the second one in terms of the expectation for the number of reads. This is proved in the next lemma.
\begin{lemma} \label{lemma:e2geqe1}
    For any $1\leq k\leq n, m=2$. We have that, $\Tmax{\cC_2}{1}\geq \Tmax{\cC_1}{1}.$
\end{lemma}
The proof can be found in Appendix \ref{Appendix_expectence_ec2_geq_ec1}.
For each $\cC_i, i\in [3]$, we conducted a simulation comprising 10 million experiments with parameters $n=35$, $m=2$, $k=20$, and for $\cC_3$, $s=2$; see Fig.~\ref{fig:sim}. We assessed the values of $\Tmax{\cC_i}{1}$. Furthermore, we assess the probability distribution of the schemes, considering them to be normally distributed based on prior research \cite{b1878965-730b-33c7-b1ad-dfd3acb6f61b},\cite{216920b6-7e32-3767-8d02-cac746ef1295} that has demonstrated this tendency using the Central Limit Theorem. To utilize this distribution.
\begin{remark} Variance analysis - 
Let us use the same notations of $t(i),T(b)$ as in \autoref{subsec:scheme2}, and since all three coding schemes are symmetric we will analyze $\nu_{(m,F)}(\cC)$ for given $\cC\in\{\cC_1, \cC_2, \cC_3\}$, and $F\subseteq[m]$.
Recall that $t(i)$ is geometrically distributed thus the variance of $t(i)$ is $Var(t(i))=\frac{mni}{(mn-i)^2}$. As each $t(i)$ is independent and identically distributed (i.i.d), $Var(T(b))=Var(\sum_{i=0}^{b-1} t(i))=\sum_{i=0}^{b-1} Var(t(i))$. By applying the Variance definition and leveraging our earlier analysis of $\E[T(b)]$ across all schemes and scenarios, we obtain the relationship: $\E[T(b)^2]=Var(T(b))+\E[T(b)]^2$.
Utilizing this formula facilitates the straightforward computation of $\E[(\nu_{(m,F)}(\cC))^2]$ by substituting the latter equation into the expression where $\E[T_b]$ is present in the $\E[\nu_{(m,F)}(\cC)]$ formula. To derive $Var(\nu_{(m,F)}(\cC))$, we again employ the definition of Variance, incorporating the previously computed values: $Var(\nu_{(m,F)}(\cC))=\E[(\nu_{(m,F)}(\cC))^2]-\E[\nu_{(m,F)}(\cC)]^2$.
The standard deviation (std) is a direct deviation from the variance calculation, expressed as $std(\nu_{(m,F)}(\cC))=\sqrt{Var(\nu_{(m,F)}(\cC))}$.
\end{remark}
Consequently, employing the normal distribution, we determine the minimum sample size required to ensure confidence levels of 90\%, 95\%, and 99\% for each coding scheme.

\begin{remark} Confidence level - Given $\cC\in\{\cC_1, \cC_2, \cC_3\}$, to evaluate the number of samples $r$ that ensure confidence level of $1-\epsilon$ by employing the normal distribution, we will use the following formula:
\begin{align*}
    r= \E[\nu_{(m,F)}(\cC)]+z\cdot std(\nu_{(m,F)}(\cC))
\end{align*}
where $z$ is the $z$-score corresponding to the desired confidence level $1-\epsilon$
obtained from standard normal distribution tables. for example for $z(1-\epsilon=0.95)=1.96$ .This approximation becomes better as $k$ or $n$ increases.
\end{remark}
Although $\Tmax{\cC_1}{1}$ might have the smallest value, $\Tmax{\cC_2}{1}$ and $\Tmax{\cC_3}{1}$ demonstrate greater stability. Notably, the number of samples ensuring a 95\% successful file recovery significantly differs from $\Tmax{\cC_1}{1}$; however, it closely aligns with $\Tmax{\cC_2}{1}$ and $\Tmax{\cC_3}{1}$. While $\mathcal{C}_3$ mirrors $\mathcal{C}_2$ in terms of expectation with minor discrepancies, $\mathcal{C}_2$ exhibits superior stability, albeit not as pronounced as $\mathcal{C}_1$. The average values of the expected number of samples for file recovery across all three coding schemes exhibited negligible disparities compared to their respective expected value  
as analyzed in \autoref{sec:random} all registering at $0.00$. Similarly, for the minimum sample sizes required to attain confidence levels of 90\%, 95\%, and 99\%, differences between the suggested normal distribution and experimental simulations were minor.

Tables \ref{tab:tab1} and \ref{tab:tab2} each provide a comparison between the simulated and analyzed values for the expected sample size required for file recovery and the corresponding confidence levels, respectively.

\begin{figure}[ht]
        \includegraphics[width=1\linewidth, height=2in]{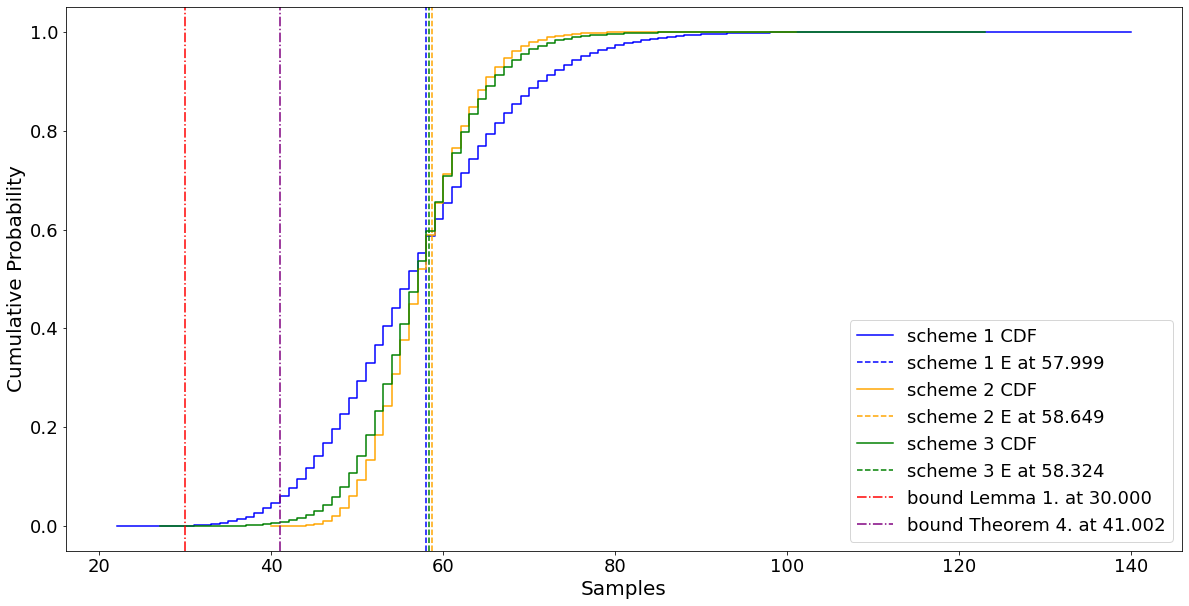}
        \caption{Illustrates the cumulative distribution function (CDF) of necessary sample sizes for file recovery and the lower bounds specified in \autoref{th: ramdom access: better lower bound} and in \autoref{lm:lowerboundmnmk}.}
        \label{fig:cdf}
\end{figure}

\begin{table}[htbp]
\centering
    \begin{tabular}{|m{2cm}|m{2cm}|m{2cm}|} 
        \hline
        coding scheme $\cC$ & avg sample size & $\Tmax{\cC}{1}$ \\
        \hline
        local MDS $\cC_1$ & 57.998 & 57.998 \\
        global MDS $\cC_2$ & 58.650 & 58.649 \\
        PMDS $\cC_3$ & 58.322 & 58.323 \\
        \hline
    \end{tabular}
    \caption{Comparison between simulation average sample size value and the expected number as analyzed in \autoref{sec:random}. }
    \label{tab:tab1}
\end{table}
\begin{table}[htbp]
\centering
    \begin{tabular}{|m{1.5cm}|m{1.5cm}|m{1.5cm}|m{1.5cm}|} 
        \hline
        coding scheme $\cC$ & confidence level & simulation & distribution\\
        \hline
        local MDS $\cC_1$ & 90\%\newline 95\%\newline99\% & 72\newline77\newline88 & 76.019 \newline 79.470 \newline 86.218\\\hline
        global MDS $\cC_2$ & 90\%\newline95\%\newline99\% & 66\newline69\newline74 & 68.026 \newline 69.821 \newline 73.333\\\hline
        PMDS $\cC_3$ & 90\%\newline95\%\newline99\% & 67\newline70\newline77 & 69.530 \newline 71.676 \newline 75.872 \\
        \hline
    \end{tabular}
    \caption{Comparison between simulation confidence level and the approximation from the normal distribution as described in \autoref{sec:compare}.}
    \label{tab:tab2}
\end{table}
In Figure \ref{fig:cdf}, the cumulative distribution function (CDF) is presented, aligning with our assumptions and evaluations. It illustrates that in terms of expectation, the local scheme $\mathcal{C}_1$ outperforms the others. This conclusion is drawn from the graph, where the blue line representing this scheme consistently surpasses the others until the sample size matches its expected number of samples needed for file recovery. As anticipated, it ensures a 0.5 confidence level that the required sample value is lower. Beyond this point, both the global and PMDS schemes ($\mathcal{C}_2$ and $\mathcal{C}_3$ respectively) demonstrate greater stability, ensuring that for all confidence levels exceeding 60\%, the sample size is smaller than $\cC_1$. Additionally, minor differences between $\mathcal{C}_1$ and $\mathcal{C}_2$ are noticeable.

Considering this comprehensive analysis, the coding scheme $\cC_2$ emerges as a preferable choice. In essence, while expected values are crucial, factors like stability also wield significant influence. Thus, identifying the optimal scheme demands meticulous analysis and consideration of various factors, pivotal for achieving the overarching goal of minimization.

\section{Random Access Expectation for Multiple Files}\label{sec:analysisageq1}
In this section, we extend the results in the paper to randomly accessing multiple files, i.e., $a\hspace{-1ex}>\hspace{-1ex}1$. We will analyze the expectation results of the first two coding schemes and show how to extend the bound from \autoref{th: ramdom access: better lower bound}. For both analyses, let us use the same notations of $t(i), T(b)$ and the Markov chain properties as mentioned in \autoref{subsec:scheme2}.
We start with the local MDS scheme. We let $F\subseteq[m], \abs{F}=a$ be the set of requested files. Assume without loss of generality that $F=\{1,\ldots,a\}$ Then, the Markov chain is described as follows.
\begin{itemize}
    \item \textbf{States definition:} The set of states is: \begin{small}
        $\cS_3=\set{\bff=(f_1,\ldots,f_{a},f_{a+1})}{\forall i\in F, f_i\leq n, f_{a+1}\leq mn-an}$.
    \end{small} where for each $i$ in $F$, $f_i $ is the number of strands drawn from the $n$ encoded strands of file $i$ and $f_{a+1}$ is the number of strands that were drawn from the other files (i.e., the other $mn-an$ strands). Given state $\bff$, let $(\bff_{-i},b)$ denote a new state where only the $i$-th value of $\bff$ changes to $b$, that is, $(\bff_{-i},b)\triangleq(f_1,f_2,\ldots,b,\ldots, f_{a+1})$.
    \item \textbf{Transition matrix:} 
    The valid transitions in $M$ (i.e., $M_{\bff,\bfy}\neq 0$) and their values are:
    \vspace{-1ex}
    \begin{small}
        \begin{align*}
        &M_{\bff,(\bff_{-i},f_i +1) } = \frac{n-f_i}{mn-\widehat{\bff}},  1 \leq i \leq a,\\
        &M_{\bff,(\bff_{-(a+1)},f_{a+1} +1)} = \frac{n\cdot(m-a) - f_{a+1}}{mn-\widehat{\bff}}, i=a+1.
    \end{align*}
    \end{small}

The next claim provides a closed formula for $M_{s_0,\bff}^{\widehat{\bff}}$, which holds for the non-absorbing states. 
    \begin{claim} 
    \label{claim:prg:C1 a geq 1}        $M_{s_0,\bff}^{\widehat{\bff}}=\frac{\binom{n}{f_1}\binom{n}{f_2}\binom{n}{f_3}\dots \binom{n}{f_{a}}\binom{n(m-a)}{f_{a+1}}}{\binom{mn}{\widehat{\bff}}}$     
    \end{claim}
    \vspace{-1.5ex}
    \begin{proof}
     At state $\bff$, we have $\binom{n}{f_i}$ options to choose $f_i$ encoded strands of file $i$ for each $i$ in $F$ and $\binom{n(m-a)}{f_{a+1}}$ options to choose $f_{a+1}$ strands from the rest of the strands in the pool, considering all possibilities for drawing a total of $\widehat{\bff}$ strands out of the $mn$ strands, which is $\binom{mn}{\widehat{\bff}}$.

    \end{proof}
  \vspace{-2ex}
        \item \textbf{Absorbing states:} 
        These are the states that allow to recover the files in $F$, so the drawing process ends. For  $\cC_1$, the absorbing states are those where we drew the $k$-th strand from file $j \in F $, which is last to be recovered, i.e., we already read at least $k$ strands from the other files in $F$ and $j$ is the last one. Denote $\Theta_1$ as the set of absorbing states. Our approach involves investigating the transient states prior to absorption since those states determine a specific absorbing state. Denote $\cG_1$ as the set of states reachable from a non-absorbing state to an absorbing one.
        \vspace{-4ex}
    \begin{align*}
        \cG_1\triangleq\set{\bff}{\exists j \in F \forall i \in F\setminus\{j\} (f_i\geq k \text{ and } {f_j=k-1} )}.
    \end{align*}
    For $g\in \cG_1$ with $j$ as the last file to be recovered, it is possible to reach exactly 1 absorbing state $\theta\hspace{-0.5ex}=\hspace{-0.5ex}(g_{-j},k)\hspace{-0.5ex}\in\hspace{-0.5ex}\Theta_1$. Thus the probability of reaching $\theta$ from $g$ is:
    $M_{s_0,g}^{(\widehat{g})}\cdot M_{g,\theta}$,
    which follows from the definition of $\cG_1.$
    


    
\end{itemize}
\begin{theorem}\label{lem:random:single:C1 code a geq 1}
For any $1\leq k\leq n$ and $ 1\leq a\leq m$, it holds that
\vspace{-2ex}
\begin{small}
    \begin{align*}
        &T(n,k;m,a)\leq\Tmax{\cC_1}{a} \\&=\sum_{j=1}^{a}\sum_{g\in \cG_1, g_j=k-1 } \frac{\binom{n}{g_1}\binom{n}{g_2}\dots\binom{n}{g_j}\dots \binom{n}{g_{a}}\binom{n(m-a)}{g_{a+1}}}{\binom{mn}{\widehat{g}}}\\&\cdot M_{g,(g_{-j},k) }\cdot mn(H_{mn}-H_{mn-(\widehat{g}+1)}).
    \end{align*}
\end{small}

\end{theorem} 

\begin{proof}
\hspace{-0.5ex}
Assume without loss of generality that $F=\{1,\ldots,a\}$.
    We wish to find $\E[\nu_{(m,F)}(\cC_1)]$.
    We let $Y$ be the random variable representing in which absorbing state the collection process ends.
    The expectation $\E[\nu_{(m,F)}(\cC_1)]$ is conditioned on $Y$. Hence,
    

    \vspace{-2ex}
    \begin{small}
            \begin{align*}
            &\E[\nu_{(m,F)}(\cC_1)]=\E_{Y}[\E[\nu_{(m,F)}(\cC_1)|Y]]\\&=\sum_{\theta\in \Theta_1 } \text{Pr}(Y=\theta)\cdot \E[\nu_{(m,F)}(\cC_1)|Y=\theta]\\&=^{(*)}\sum_{j=1}^{a}\sum_{g\in \cG_1, g_j=k-1 } M_{s_0,g}^{(\widehat{g})} \cdot M_{g,g_{-j}(k)}\cdot \E[T(\widehat{g}+1)]\\&=
            \sum_{j=1}^{a}\sum_{g\in \cG_1, g_j=k-1 } \frac{\binom{n}{g_1}\binom{n}{g_2}\dots\binom{n}{g_j}\dots \binom{n}{g_{a}}\binom{n(m-a)}{g_{a+1}}}{\binom{mn}{\widehat{g}}}\\&\cdot M_{g,(g_{-j},k) }\cdot mn(H_{mn}-H_{mn-(\widehat{g}+1)}),
    \end{align*}
    \end{small}
    \hspace{-2ex}where $(*)$ follows from the definitions of $\cG_1$ so we iterate over all absorbing states $\sum_{\theta\in \Theta_1}\text{Pr}(Y=\theta) = \sum_{j=1}^{a}\sum_{g\in \cG_1, g_j=k-1 }M_{s_0,g}^{(\widehat{g})}\cdot M_{g,(g_{-j},k)}$.
    Since the code is symmetric it implies that $\E[\nu_{(m,F)}(\cC_1)]=\Tmax{\cC_1}{a}.$
\vspace{-0.2ex}
\end{proof}
The next theorem states the extension result of the global MDS scheme for accessing multiple files.
\begin{theorem}
For any $1\leq k\leq n$ and $ 1\leq a\leq m$, it holds that

    \vspace{-2ex}
    \begin{small}
    \begin{align*}
        &T(n,k;m,a)\leq\Tmax{\cC_2}{a}=mn\left(H_{mn}\hspace{-0.4ex}-\hspace{-0.4ex}H_{mn-mk}\right)\\&-\hspace{-0.5ex} \frac{mn}{\binom{mn}{ak}}\Bigg(\hspace{-0.5ex}\sum_{j=0}^{mk-ak-1}\hspace{-1ex} \binom{ak\hspace{-0.5ex}-\hspace{-0.5ex}1\hspace{-0.5ex}+\hspace{-0.5ex}j}{ak\hspace{-0.3ex}-\hspace{-0.3ex}1} H_{mn-(ak+j)} \hspace{-0.5ex}-\hspace{-0.5ex} H_{mn-mk} \binom{mk\hspace{-0.5ex}-\hspace{-0.5ex}1}{ak}\hspace{-0.5ex}\Bigg).
    \end{align*}
\end{small}
\end{theorem}
\begin{proof}
Assume without loss of generality that $F=\{1,\ldots,a\}$. 
\begin{itemize}
    \item \textbf{States definition:} The set of states is $\cS_4=\set{(i,j)}{0\leq i\leq ak, 0\leq j\leq mn-ak}$, where $i$ is the number of strands drawn from the $ak$ systematic strands of the files in $F$, and $j$ is the number of strands that were drawn from the other $mn-ak$ strands.
    \item \textbf{Transition matrix:}  
    The valid transitions in $M$ (i.e. $M_{(i_1,j_1),(i_2,j_2)}\neq 0$) and their values are:
    \begin{align*}
        &M_{(i,j),(i+1,j)} = \frac{ak-i}{mn-(i+j)},
        M_{(i,j),(i,j+1)} = \frac{mn-ak-j}{mn-(i+j)}
    \end{align*}
    The next claim provides a closed formula for $M_{(0,0),(i,j)}^{(i+j)}$, which holds for the non-absorbing states.
    \begin{claim} 
    \label{claim:prg:C2 a geq 1}
        $M_{(0,0),(i,j)}^{(i+j)}=\frac{\binom{i+j}{i} \binom{mn-(i+j)}{ak-i}}{\binom{mn}{ak}}$   
    \end{claim}
    \begin{proof}
    
    At the state $(i,j)$, we have $\binom{ak}{i}$ options to choose $i$ systematic strands and $\binom{mn-ak}{j}$ options to choose $j$ strands from the rest of the pool, considering all possibilities for drawing a total of $(i+j)$ strands out of the $mn$ strands, which is $\binom{mn}{i+j}$.
    \begin{align*}
        M_{(0,0),(i,j)}^{(i+j)} &=\frac{\binom{ak}{i} \binom{mn-ak}{j}}{\binom{mn}{i+j}} = \frac{\frac{(ak)! (mn-ak)!}{(ak-i)! i! (mn-ak-j)! j!}}{\frac{(mn)!}{(i+j)! (mn-(i+j))!}} \\&=\frac{\binom{i+j}{i} \binom{mn-(i+j)}{ak-i}}{\binom{mn}{ak}}
    \end{align*}
    \end{proof}
    \item \textbf{Absorbing states:} 
    These are the states that allow to recover the files in $F$, so the drawing process ends. In scheme code $\cC_2$ the absorbing states are those where we either drew the $ak$ systematic strands of the files in $F$ or any $mk$ different strands from the pool of $mn$ strands. We denote $\Theta_1, \Theta_2$ as the set of absorbing states corresponding to the first,second option, respectively. That is,
    \begin{align*}
        &\Theta_1\triangleq\set{(ak,j)}{0\leq j\leq mk-ak-1},\\ &\Theta_2\triangleq\set{(i,mk-i)}{0\leq i\leq ak}.
    \end{align*}
    Our approach involves investigating the transient states prior to absorption since those states determine a specific absorbing state. Denote $\cG_1, \cG_2$ the set of states reachable from a non-absorbing state to an absorbing one corresponding to $\Theta_1, \Theta_2$.
    \begin{align*}
        &\cG_1\triangleq\set{(ak-1,j)}{0\leq j\leq mk-ak-1},\\ &\cG_2\triangleq\set{(i,mk-i-1)}{0\leq i\leq ak-1}.
    \end{align*}
    
    Note that:
    \begin{enumerate}
        \item Given $g=(ak-1,j)\in \cG_1$  there exists 1 corresponding absorbing state $\theta = (ak,j) \in \Theta_1$, (since $(ak,j-1)$ is also an absorbing state), thus the probability of reaching the absorbing state $\theta$ is, 
        $$M_{(0,0),g}^{(\widehat{g})}\cdot M_{g,\theta}.$$
    \item Given $g=(i,mk-i-1)\in \cG_1$  there exists 2 corresponding absorbing states $\theta_1 = (i,mk-i), \theta_2 =  (i+1,mk-i-1) \in \Theta_2$, thus the probability of reaching the absorbing states $\theta_1, \theta_2$ is, 
    \begin{align*}
        M_{(0,0),g}^{(\widehat{g})}\cdot (M_{g,\theta_1} +M_{g,\theta_2})=M_{(0,0),g}^{(\widehat{g})}.
    \end{align*}
    \end{enumerate}

    \item \textbf{The expectation:}
    In order to calculate $\E[\nu_{(m,F)}(\cC_2)]$, we let $Y$ be the random variable representing in which absorbing state the collection process ends. The expectation $\E[\nu_{(m,F)}(\cC_2)]$ is conditioned on $Y$. Hence,
    \begin{small}
        \begin{align}
        \nonumber&\E[\nu_{(m,F)}(\cC_2)]=\E_{Y}[\E[\nu_{(m,F)}(\cC_2)|Y]]\\&\nonumber=\sum_{\theta\in \Theta_1\cup \Theta_2 } \text{Pr}(Y=\theta)\cdot \E[\nu_{(m,F)}(\cC_2)|Y=\theta]\\&\nonumber=\sum_{\theta\in \Theta_1\cup \Theta_2 } M_{(0,0),\theta}^{(\widehat{\theta})}\cdot \E[T(\widehat{\theta})]=mn\left(H_{mn}\hspace{-0.4ex}-\hspace{-0.4ex}H_{mn-mk}\right)\\&\nonumber-\hspace{-0.5ex} \frac{mn}{\binom{mn}{ak}}\Bigg(\hspace{-0.5ex}\sum_{j=0}^{mk-ak-1}\hspace{-1ex} \binom{ak\hspace{-0.5ex}-\hspace{-0.5ex}1\hspace{-0.5ex}+\hspace{-0.5ex}j}{ak\hspace{-0.3ex}-\hspace{-0.3ex}1} H_{mn-(ak+j)} \hspace{-0.5ex}\\&-\hspace{-0.5ex} H_{mn-mk} \binom{mk\hspace{-0.5ex}-\hspace{-0.5ex}1}{ak}\hspace{-0.5ex}\Bigg). \label{eq:excpc2a geq 1}
    \end{align}
    \end{small}
    
    The full proof of (\ref{eq:excpc2a geq 1}) can be found in Appendix \ref{Appendix_expectence_c2}.
    Since the code is symmetric it implies that
    $$\E[\nu_{(m,F)}(\cC_2)]=\Tmax{\cC_2}{a}.$$
    
\end{itemize}
\end{proof}

Lastly, we present our lower bound of $T(n,k;m,a)$.
\begin{theorem}\label{th: ramdom access: improved lower bound}
    Let $\cC$ be an $[mn,mk]$ code. It holds that 

    \begin{small}
        \begin{align*}
         &T(n,k;m,a) \ge \frac{mn}{\binom{m}{a}}\sum_{i=a-1}^{m-1}  \binom{i\hspace{-0.5ex}-\hspace{-0.5ex}1}{a\hspace{-0.5ex}-\hspace{-0.5ex}1}(m\hspace{-0.5ex}-\hspace{-0.5ex}i)(H_{mn-ki}\hspace{-0.5ex}-\hspace{-0.5ex}H_{mn-k(i+1)}).
    \end{align*}
    \end{small}
\end{theorem}
\begin{proof}
Let us use the same notations
as in the proof of \autoref{th: ramdom access: better lower bound}. Since the $i$-th file that is recovered can belong to any subset of files with size $a$ that were all recovered before this file i.e., it is the last to be recovered, we count this time for all $\binom{i-1}{a-1}$ sets of $a$ requested files. We have that
\begin{align*}
   \sum_{J\subseteq[m], \abs{J}=a} \nu_{(m,F)}(\cC) = \sum_{i=1}^m \binom{i-1}{a-1}n_i(v)=\sum_{i=a}^m \binom{i-1}{a-1}n_i(v) 
\end{align*}
Hence, 
    \begin{align*}
        &\sum_{J\subseteq[m], \abs{J}=a} \E[\nu_{(m,F)}(\cC)] = \E\left[\sum_{J\subseteq[m], \abs{J}=a} \nu_{(m,F)}(\cC)\right]  \\&\ge\E\left[\sum_{i=a}^m \binom{i-1}{a-1}\sum_{j=1}^{ki} t_j(\cC)\right] = \sum_{i=a}^m\binom{i-1}{a-1}\sum_{j=1}^{ki} \E\left[t_j(\cC)\right].
    \end{align*}
Note that for any $j\in[mk]$, we have that $t_j(\cC)$ is a geometric random variable with success probability $p_j=\frac{mn-(j-1)}{mn}$  and so $\E[t_j(\cC)] = \frac{mn}{mn-(j-1)}$, and
    \begin{align*}
    &\sum_{J\subseteq[m], \abs{J}=a} \E[\nu_{(m,F)}(\cC)]  \ge  \sum_{i=a}^m\binom{i-1}{a-1}\sum_{j=1}^{ki} \E\left[t_j(\cC)\right] 
    \\&= \sum_{i=a}^m \binom{i-1}{a-1}\sum_{j=1}^{ki} \frac{mn}{mn-(j-1)}\\
    & = mn \sum_{i=a}^m \binom{i-1}{a-1}\left(\frac{1}{mn-ki+1} + \frac{1}{mn-ki+2} + \ldots + \frac{1}{mn}\right) \\
    &= mn \sum_{i=a-1}^{m-1}  \binom{i-1}{a-1}(m-i) \left( \sum_{j=0}^{k-1}\frac{1}{mn-ki -j}\right)\\
    &=mn \sum_{i=a-1}^{m-1}  \binom{i-1}{a-1}(m-i)(H_{mn-ki}-H_{mn-k(i+1)}).
    \end{align*}
    Hence we have that 
    \begin{align*}
    &\frac{1}{\binom{m}{a}}  \sum_{J\subseteq[m], \abs{J}=a} \E[\nu_{(m,F)}(\cC)] \\&\geq\frac{mn}{\binom{m}{a}}\sum_{i=a-1}^{m-1}  \binom{i-1}{a-1}(m-i)(H_{mn-ki}-H_{mn-k(i+1)}).
    \end{align*}
    In particular, there exists $J\subseteq[m], \abs{J}=a$ for which $\E[\nu_{(m,F)}(\cC)]  \geq \frac{mn}{\binom{m}{a}}\sum_{i=a-1}^{m-1}  \binom{i-1}{a-1}(m-i)(H_{mn-ki}-H_{mn-k(i+1)}).$, i.e., $\Tmax{\cC}{a}\geq T(n,k;m,a)\geq \frac{mn}{\binom{m}{a}}\sum_{i=a-1}^{m-1}  \binom{i-1}{a-1}(m-i)(H_{mn-ki}-H_{mn-k(i+1)}).$ 
\end{proof}
\section{Conclusion And Future Work}
This paper investigates the random access coverage depth problem in practical scenarios, focusing on storing $m$ files and retrieving portions of them. By analyzing the maximal expected number of samples required for file recovery and $T(n,k;m,a)$, the study sheds light on the structural attributes of various coding schemes that impact random access expectations and probability distributions. While the findings represent significant progress in this domain, several intriguing avenues for future research remain unexplored. In our future research we will  extend our analysis to encompass the more general setup of comparing across all 3 schemes when $a,m \geq 2$, and plan to find the exact value of $ T(n,k;m,a)$ and study the probability distribution additionally, attention will be directed towards addressing challenges related to the noisy channel of DNA storage, specifically concerning \autoref{prob:random:single} and \autoref{prob:optimal}.

\section{Acknowledgment}
The authors wish to thank Zohar Nagel for her progress on initial results on the problems studied in the paper. They also thank Ido Feldman for helpful discussions.

\newpage
\bibliographystyle{ieeetr}
\bibliography{bib}
\newpage
\appendices
\newpage

\section{}
\label{Appendix_expectence_c2}
\subsection*{Proof of $\E[\nu_{(m,F)}(\cC_2)]$ for multiple files}
\begin{small}
    \begin{align*}
    &\E[\nu_{(m,F)}(\cC_2)]=\E_{Y}[\E[\nu_{(m,F)}(\cC_2)|Y]]\\&=\sum_{\theta\in \Theta_1\cup \Theta_2 } \text{Pr}(Y=\theta)\cdot \E[\nu_{(m,F)}(\cC_2)|Y=\theta]\\&=\sum_{\theta\in \Theta_1\cup \Theta_2 } M_{(0,0),\theta}^{(\widehat{\theta})}\cdot \E[T(\widehat{\theta})]\\&
    =\sum_{\theta\in \Theta_1} M_{(0,0),\theta}^{(\widehat{\theta})}\cdot \E[T(\widehat{\theta})]+\sum_{\theta\in \Theta_2} M_{(0,0),\theta}^{(\widehat{\theta})}\cdot \E[T(\widehat{\theta})]\\&
    =\sum_{g=(ak-1,j)\in \cG_1} M_{(0,0),(ak-1,j)}^{(ak-1+j)}\cdot M_{(ak-1,j),(ak,j)}\E[T(ak+j)]\\&+\sum_{g=(i,mk-i-1)\in \cG_2} M_{(0,0),(i,mk-i-1)}^{(mk-1)}\cdot \E[T(mk)]\\
    &=\sum_{j=0}^{mk-ak-1} \left(\frac{\binom{ak-1+j}{ak-1}}{\binom{mn}{ak}} \cdot mn \left(H_{mn} - H_{mn-(ak+j)}\right)\right) \\&+ mn \left(H_{mn} - H_{mn-mk}\right)\cdot \sum_{i=0}^{ak-1} \frac{\binom{mk-1}{i} \cdot \binom{mn-(mk-1)}{ak-i}}{\binom{mn}{ak}}\\
    &=\frac{mn}{\binom{mn}{ak}} \Bigg(\sum_{j=0}^{mk-ak-1} \binom{ak-1+j}{ak-1} \cdot \left(H_{mn} - H_{mn-(ak+j)}\right) \\&+ \left(H_{mn} - H_{mn-mk}\right) \sum_{i=0}^{ak-1} \binom{mk-1}{i}\binom{mn-(mk-1)}{ak-i}\Bigg)\\
    &=\frac{mn}{\binom{mn}{ak}} \Bigg(\sum_{j=0}^{mk-ak-1} \binom{ak-1+j}{ak-1} \cdot \left(H_{mn} - H_{mn-(ak+j)}\right) \\&+ \left(H_{mn} - H_{mn-mk}\right) \left(\binom{mn}{ak} - \binom{mk-1}{ak}\right)\Bigg)\\
    &=\frac{mn}{\binom{mn}{ak}} \Bigg(H_{mn} \binom{mk-1}{ak} - \sum_{j=0}^{mk-ak-1} \binom{ak-1+j}{ak-1}H_{mn-(ak+j)} \\&+ \left(H_{mn} - H_{mn-mk}\right) \left(\binom{mn}{ak} - \binom{mk-1}{ak}\right)\Bigg)\\
    &=\frac{mn}{\binom{mn}{ak}} \Bigg(H_{mn} \left(\binom{mk-1}{ak}+\binom{mn}{ak}-\binom{mk-1}{ak}\right)\\&-\sum_{j=0}^{mk-ak-1} \binom{ak-1+j}{ak-1}H_{mn-(ak+j)} \\&- H_{mn-mk}\left(\binom{mn}{ak} - \binom{mk-1}{ak}\right)\Bigg)\\
    &=^{(*)}\frac{mn}{\binom{mn}{ak}} \Bigg(H_{mn}\binom{mn}{ak}-\sum_{j=0}^{mk-ak-1}\binom{ak-1+j}{ak-1}H_{mn-(ak+j)}\\&-H_{mn-mk}\left(\binom{mn}{ak} - \binom{mk-1}{ak}\right)\Bigg)\\
    &=mn(H_{mn}-H_{mn-mk})\\&-\frac{mn}{\binom{mn}{ak}} \Bigg(\sum_{j=0}^{mk-ak-1} \binom{ak-1+j}{ak-1}H_{mn-(ak+j)} \\&-H_{mn-mk}\binom{mk-1}{ak}\Bigg),
    \end{align*}
\end{small}

    where $(*)$ follows from Pascal identity.

\section{}
\label{Appendix_expectence_ec2_geq_ec1}
\subsection*{Proof of \autoref{lemma:e2geqe1}}
\begin{proof}
\begin{small}
  \begin{align*}
    &\frac{\Tmax{\cC_2}{1}-\Tmax{\cC_1}{1}}{2n}=\left( H_{2n} - H_{2n-2k} \right) - \left( H_{n} - H_{n-k} \right)\\&- \frac{1}{{2n \choose k}} \left( \sum_{j=0}^{k-1} \binom{k-1+j}{k-1} \cdot H_{2n-(k+j)} - H_{2n-2k} \cdot \binom{2k-1}{k} \right)
\end{align*}  
\end{small}

    In the subsequent analysis, we simplify the equation by decomposing it into two sub-equations. As for the first part:
    \begin{small}
        \begin{align*}
        &\left( H_{2n}-H_{2n-2k}\right)-\left( H_{n}-H_{n-k}\right)=
        \sum_{i=2n-2k+1}^{2n}\frac{1}{i} - \sum_{i=n-k+1}^{n}\frac{1}{i}\\&=
        \sum_{i=n-k+1}^{n}\left(\frac{1}{2i-1}+\frac{1}{2i}\right)-\sum_{i=n-k+1}^{n}\frac{1}{i}\\&=^{(*)}
        \sum_{i=n-k+1}^{n}\left(\frac{1}{i}+\frac{1}{2i(2i-1)}\right)-\sum_{i=n-k+1}^{n}\frac{1}{i}\\&=
        \sum_{i=n-k+1}^{n}\left(\frac{1}{2i(2i-1)}\right)\\&=
        \sum_{t=0}^{k-1}\frac{1}{(2n-2k+2t+2)(2n-2k+2t+1)}
        \end{align*}
    \end{small}
        
    where $(*)$ is $\frac{1}{2i - 1} = \frac{1}{2i} + \frac{1}{2i \cdot (2i - 1)}.$\\
    As for the second part:
    \begin{small}
        \begin{align*}
            &\frac{1}{\binom{2n}{k}} \left(\sum_{j=0}^{k-1} \binom{k-1+j}{k-1} \cdot H_{2n-(k+j)} - H_{2n-2k} \binom{2k-1}{k} \right)\\&
            =\frac{1}{\binom{2n}{k}} \left( \sum_{j=0}^{k-1} \binom{k-1+j}{k-1}(H_{2n-(k+j)} - H_{2n-2k}) \right)\\&
            =\frac{1}{\binom{2n}{k}} \left(\sum_{j=0}^{k-1} \binom{k-1+j}{k-1}\left( \sum_{i=2n-2k+1}^{2n-(k+j)} \frac{1}{i} \right) \right)\\
            &=^{(**)}\frac{1}{\binom{2n}{k}}\left(\sum_{t=0}^{k-1} \frac{1}{(2n-2k+t+1)} \binom{2k-1-t}{k} \right)
        \end{align*}
    \end{small}
     where $(**)$ follows Pascal identity $\sum_{j=0}^{k-1-t} \binom{k-1+j}{k-1} = \binom{2k-1-t}{k}$ .

    combining all together we get:
    \begin{small}
        \begin{align}
            \nonumber&\sum_{t=0}^{k-1} \left( \frac{1}{(2n-2k+2t+2)(2n-2k+2t+1)} \right)\\&\nonumber-\frac{1}{\binom{2n}{k}}\sum_{t=0}^{k-1}\left(\frac{1}{(2n-2k+t+1)}\binom{2k-1-t}{k} \right)\\&
            \nonumber=\sum_{t=0}^{k-1} \Bigg( \frac{1}{(2n-2k+2t+2)(2n-2k+2t+1)}\\&-\frac{1}{(2n-2k+t+1)}\cdot \frac{\binom{2k-1-t}{k}}{\binom{2n}{k}} \Bigg)\label{eq:diffe1e2}
        \end{align}
    \end{small}
    
    Let us examine each element in the sum,
    \begin{small}
        \begin{align*}
            &g(t)\triangleq \frac{1}{(2n-2k+2t+2)(2n-2k+2t+1)}\\&-\frac{1}{(2n-2k+t+1)} \cdot \frac{\binom{2k-1-t}{k}}{\binom{2n)}{k}}.
        \end{align*}
    \end{small}
    The next claim proves that $g(t)$ is a decreasing function of $t$.
    \begin{claim}\label{claim:decreasing g(t)}
        For $1\leq k \leq n$ and $0\leq t\leq k-2$ it holds that: $g(t+1)< g(t)$.
    \end{claim}
    \begin{proof}
    We will demonstrate that $g(t+1)< g(t)$ by establishing $g(t)- g(t+1)> 0$, we simplify the equation by decomposing it into two sub-equations.As for the first part:
    \begin{small}
        \begin{align*}
            &\frac{1}{(2n-2k+2t+2)(2n-2k+2t+1)}\\&-\frac{1}{(2n-2k+2(t+1)+2)(2n-2k+2(t+1)+1)}>0
        \end{align*}
    \end{small}
    Directly.\\
    As for the second part:
    \begin{small}
        \begin{align*}
            &\frac{1}{(2n-2k+(t+1)+1)} \cdot \frac{\binom{2k-t-2}{k}}{\binom{2n}{k}} -\frac{1}{(2n-2k+t+1)} \cdot \frac{\binom{2k-t-1}{k}}{\binom{2n}{k}}\\&
            =\frac{\binom{2k-t-2}{k}}{\binom{2n}{k}}\left(\frac{1}{(2n-2k+t+2)} - \frac{2k-t-1}{(2n-2k+t+1)\cdot(k-t-1)} \right)\\&
            =\frac{\binom{2k-t-2}{k}}{\binom{2n}{k}(2n-2k+t+1)(k-t-1)(2n-2k+t+2)}\\&\cdot \left( t(1-k)+2nk+2k^2-3k+1\right).
        \end{align*}
    \end{small}
    
    Let,
    \begin{small}
        \begin{align*}
            &h(t) \triangleq \frac{\binom{2k-t-2}{k}}{\binom{2n}{k}(2n-2k+t+1)(k-t-1)(2n-2k+t+2)}\\&\cdot(t(1-k)+2nk+2k^2-3k+1),
            \\&H_1(t) \triangleq\frac{\binom{2k-t-2}{k}}{\binom{2n}{k}(2n-2k+t+1)(k-t-1)(2n-2k+t+2)},\\&
            H_2(t)\triangleq t(1-k)+2nk+2k^2-3k+1.
        \end{align*}
    \end{small}
   
        $H_1(t)>0$ directly.
        For $k=1$ $H_2(t)$ is a positive function thus $h(t)=H_1(t)\cdot H_2(t)=H_1(t) \cdot 2n>0$\\
        for $2\leq k \leq n$ $H_2(t)$ is a decreasing linear function with respect to $t$ due to the negative coefficient $1-k<0$. Next, our objective is to demonstrate that for $t=k-2$ the function evaluates to a positive value, thereby establishing the desired conclusion.
        \begin{small}
            \begin{align*}
                &h(k-2) = H_1(t) \cdot \left( (k-2) \cdot (1-k) + 2nk + 2k^2 - 3k + 1 \right) \\&= H_1(t) \cdot (2nk + k^2 - 1) \geq H_1(t) \cdot (2k^2 + k^2 - 1) \\&= H_1(t) \cdot (3k^2 - 1) > 0
            \end{align*}
        \end{small}
    \end{proof}

    Considering \autoref{claim:decreasing g(t)} we have that: 
    $$(\ref{eq:diffe1e2})=\sum_{t=0}^{k-1} g(t) >0$$
    \end{proof}

\end{document}